\documentclass[aps,twocolumn,prd,nofootinbib,superscriptaddress]{revtex4-1}

\usepackage{geometry}
\usepackage{graphicx}
\usepackage[caption=false]{subfig}
\usepackage{mathrsfs}

\usepackage{amsmath, bm}
\usepackage{hhline}
\usepackage{xcolor}

\usepackage{verbatim}
\usepackage{dcolumn}
\usepackage{color}
\usepackage{xspace}
\usepackage{url}
\usepackage{float}
\usepackage{multirow}
\usepackage{hyperref}
\usepackage[dvipsnames]{xcolor}

\newcommand{\secref}[1]{Section~\ref{#1}}
\newcommand{\appendref}[1]{Appendix~\ref{#1}}
\newcommand{\figref}[1]{Figure~\ref{#1}}
\newcommand{\tabref}[1]{Tab.~\ref{#1}}
\newcommand{\eqnref}[1]{\eqref{#1}}                       

\newcommand{\Figref}[1]{Figure~\ref{#1}}
\newcommand{\Tabref}[1]{Table~\ref{#1}}

\newcommand\prx{Phys.~Rev.~X}
\def\apj{Astrophysical Journal}                %
\def\apjl{Astrophysical Journal}             %

\def\prd{Phys.~Rev.~D}       %
\def\prl{Phys.~Rev.~Lett}    %
\def\cqg{Class.~Quant.~Grav.~}%

\newcommand{\Lmarg}{\mathcal{L}_{\rm{marg}}(\pmb{\lambda})}
\newcommand{\lnLmarg}{\ln \mathcal{L}_{\rm{marg}}}
\newcommand{\lnL}{\ln \mathcal{L}}

\newcommand{\teo}{\texttt{TEOBResumS}}
\newcommand{\mc}{{\cal M}_{\rm{c}}}

\newcommand{\chieff}{\chi_{\rm eff}}

\newcommand\addition[1]{{\color{violet}#1}}
\newcommand\revthree[1]{{\color{black}#1}}

\newcommand\CarlosO[1]{{\color{black}#1}}

\def\RIT{Center for Computational Relativity and Gravitation, Rochester Institute of Technology, Rochester, New York 14623, USA}
\def\UNICAL{Dipartimento di Fisica, Universit\`a della Calabria, Arcavacata di Rende (CS), 87036, Italy}

\begin{document}
\title{Parameter Estimation with Targeted Eccentric Numerical-Relativity Simulations for GW200208\_22 and GW190620}
\author{Patricia McMillin}
\affiliation{\RIT}
\author{Katelyn J. Wagner}
\affiliation{\RIT}
\author{Giuseppe Ficarra}
\affiliation{\UNICAL}
\author{Carlos O. Lousto}
\affiliation{\RIT}
\author{Richard O'Shaughnessy}
\affiliation{\RIT}
\begin{abstract}
We have analyzed LVK gravitational wave events that show some evidence of 
eccentricity from TEOBResumS modeling parameter estimations and have confronted 
them independently with full numerical generated waveforms from our bank of 
nearly two thousand simulations of binary black holes. We have used RIFT for 
Bayesian parameter estimation and found that GW200208\_{22} KDE 
estimates favor eccentricities $e_{20} = 0.198_{-0.180}^{+0.119}$ 
upon entering the LVK band at $\sim20$Hz within a 90\% \revthree{confidence interval}. 
Within this event analysis we employed 42 new targeted full numerical 
relativity simulations
and we have thus found a top improved likelihood $\lnL$ matching waveform, 
compared to model-based analysis, with an estimated eccentricity at 20Hz, 
$e_{20}=0.200$,
thus reinforcing the eccentric hypothesis of the binary.
We have also used our full bank of 
numerical waveforms on GW190620 finding that the KDE estimate favors eccentricities at 10 Hz in $e_{10}=0.190_{-0.186}^{+0.046}$.
New specifically targeted simulations will be required to narrow these 
eccentricity ranges.
\end{abstract}
\maketitle
 
\section{Introduction}
Ground-based gravitational wave (GW) detectors in the International 
Gravitational Wave observatory Network (IGWN), including Advanced LIGO  
\cite{2015CQGra..32g4001L} and Virgo \cite{2015CQGra..32b4001A}, now joined by 
KAGRA \cite{2021PTEP.2021eA101A} continue to identify coalescing compact 
binaries \cite{DiscoveryPaper,LIGO-O1-BBH,LIGO-GW170817-bns,LIGO-O3-NSBH,LIGO-O3-O3a_final-catalog,LIGO-O3-O3b-catalog}.
Their properties are understood by comparing gravitational wave (GW) 
observations to estimates of the radiation emitted from merging binary black 
holes, produced by detailed numerical calculations or phenomenological 
estimates.

Most compact binaries that enter the LIGO-Virgo-KAGRA (LVK) sensitivity band 
at 10Hz are expected to have quasi-circular orbits. Systems formed through 
isolated binary evolution form at large enough separations that their merger 
timescales are sufficiently large such that gravitational radiation is expected 
to circularize their orbit long before merger 
\cite{Bethe_1998, Belczynski_2002, Stevenson_2017}. 
However,recent studies have suggested that some compact binary systems may 
instead display eccentric orbits, with some orbital eccentricity remaining 
at the time of merger. Systems with eccentric orbits may indicate an 
alternative formation channel, called dynamical assembly.

Binaries may form via this channel due to high rates of chance encounters 
in densely populated stellar environments \cite{Mandel_2010}, such as globular 
clusters \cite{Sadowski:2007dz,Rodriguez_2018a,Zevin_2019}, nuclear star 
clusters \cite{Fragione_2020}, or galactic centers \cite{Samsing_2014}, where 
multiple black hole interactions might be favored 
\cite{Campanelli:2007ea,Lousto:2007rj,Ficarra:2023zjc,Ficarra:2024jen,Bamber:2025gxj,Heinze:2025usf}. 
In these environments, objects that have already evolved into compact objects 
may encounter each other, and undergo one or a series of gravitational 
encounters that enable binary systems to form and merge on short timescales. 
These binaries are expected to be short-lived, but may merge with 
non-negligible eccentricity that is likely measurable with current and 
future gravitational wave detectors \cite{Rodriguez_2018b}.

There have been several recent studies that attempt to probe eccentricity 
in binary mergers detected by LIGO and Virgo. Romero-Shaw et al. 
(2020, 2021, 2022) \cite{RomeroShaw_2020,RomeroShaw_2021,Romero-Shaw:2022xko} 
use an efficient reweighting method \cite{Romero_Shaw_2019,Payne_2019} to 
obtain measurements of the orbital eccentricity for gravitational-wave sources 
up to and including the third LVK gravitational-wave transient catalog, GWTC-1 
\cite{LIGO-O1-BBH}, GWTC-2.1 \cite{LIGO-O3-O3a_final-catalog}, and GWTC-3 
\cite{LIGO-O3-O3b-catalog}. This novel postprocessing technique first analyzes 
the catalog with the ``proposed" spin-aligned quasi-circular waveform model 
\texttt{IMRPhenomD} \cite{Khan:2015jqa}, these posteriors are reweighted to the 
``target" model: spin-aligned eccentric waveform \texttt{SEOBNRE} 
\cite{seobnre,Liu_2020_seobnre}. Results from this 
method are unable to distinguish between spin-induced precession and 
eccentricity, and are limited to moderate eccentricities at 10 Hz 
($e_{10}\leq0.2$) with restricted spins ($\chieff \leq 0.6$). Romero-Shaw et 
al. (2022) \cite{Romero_Shaw_2022} found that GW190521, GW190620, GW191109, and 
GW200208\_22 show considerable support for moderate eccentricity 
($e_{10}\geq 0.05$) where the latter two events were new 
additions to the population. 

Gupte et al. \cite{Gupte:2024jfe} also identified GW200208\_22 
as eccentric from the first three LVK catalogs 
\cite{LIGO-O1-BBH,LIGO-O3-O3a_final-catalog,LIGO-O3-O3b-catalog}, as well 
as adding GW200129, and GW190701 to the population of events with signs of 
eccentricity. These results were obtained through Bayesian inference with two 
eccentric parameters (eccentricity and relativistic anomaly) on the LVK catalog 
using \texttt{DINGO} \cite{Dax_2021} with the spin-aligned quasi-circular model 
\texttt{SEOBNRv4HM} \cite{Bohe:2016gbl,Cotesta_2018} and the spin-aligned 
eccentric model \texttt{SEOBNRv4EHM} \cite{Ramos_Buades_2022,Khalil_2021}. 
The eccentric models used were limited to $e_{10}\leq 0.5$ and are not 
simultaneously able to include both spin-precession effects and eccentricity, 
thus are unable to distinguish between the two effects.

A study of the evolution of hierarchical triple systems done in 
\cite{Ficarra:2023zjc,Ficarra:2024jen} found that scattering of the distant 
companion on the inner binary is more effective at imparting 
eccentricity onto the binary than if the companion is in a distant 
quasi-circular orbit.
While Romero-Shaw et al. (2025) \cite{romero-shaw_2025} completed a 
comparison of the properties of GW200208\_22 as analyzed in 
\cite{Romero-Shaw:2022xko,Gupte:2024jfe} and discussed the astrophysical implication 
of this event having non-negligible eccentricity present at merger. 
They claim that
this event almost certainly formed in a dynamical environment, where an 
analysis of formation scenarios for this event shows that GW200208\_22 could 
have plausibly been formed in a hierarchical field triple or within a 
globular cluster but unlikely to have been the result of formation in an 
active galactic nucleus as it is dependent on disk geometry and the binary's location in the disk.

Iglesias et al. \cite{Iglesias_2024} use \texttt{TEOBResumS-Dalí} 
\cite{teobresums2,teobresums_general,Albanesi:2022xge}, which can 
generate stable waveforms for $e\leq 0.9$ where the $e\leq0.2$ waveforms are 
verified via comparison to numerical relativity waveforms, for Bayesian 
parameter estimation using RIFT \cite{rift}. Eccentric reanalysis on GW150914, 
GW190521, GW190620, GW190706, and GW190929 returned no strong preference for 
eccentricity for any of the events, although eccentricity can not be ruled out 
completely due to the lack of waveform model that simultaneously includes 
spin-induced precession and eccentricity. 

In Healy et al. \cite{Healy:2020jjs} using the third RIT BBH simulations 
catalog \cite{Healy:2020vre} (with $e\approx0$) and RIFT techniques applied 
to LIGO/Virgo’s O1/O2 observational runs, we obtained improved binary 
parameters, extrinsic parameters, and the remnant properties of thirteen 
gravitational waves events. And then again for the first GW event in the O3 
observational run, GW170729, we used the RIFT with NR waveforms approach to 
successfully match it in \cite{Chatziioannou:2019dsz}. Cross-check of the RIT 
numerical waveforms with the completely independent SXS NR implementation for 
the match to the event GW150914 have been performed for up to $\ell=5$ modes, 
showing a high degree of overlap and convergence toward each other's results 
in \cite{Lovelace:2016uwp}. 

Gayathri et al. (2022) \cite{Gayathri:2020coq} 
showed that GW190521 is most consistent with a highly eccentric black hole 
merger. They generated 611 targeted numerical relativity eccentric simulations, 
for an effective $\sim6\times10^4$ gravitational waveforms with different total 
masses, much greater than previously available at high eccentricities. 
These NR simulations were compared to the observed data using RIFT \cite{rift}, 
they found that GW190521 is best explained by a
high-eccentricity, precessing model with $e\sim0.7$. All properties of
GW190521 point to its origin being the repeated gravitational capture
of black holes, making GW190521 one of the first of LIGO/Virgo's discoveries
whose formation channel is identified.

Gamba et al. (2022) \cite{Gamba_2022} also analyzed GW190521 under the 
hypothesis of dynamical capture where the binary components are on hyperbolic 
orbits. They utilize \texttt{TEOBResumS} waveforms in both hyperbolic, and spin 
precessing flavors to perform Bayesian parameter estimation. Gamba et al. 
found that GW190521 favored a dynamical capture scenario with hyperbolic 
orbits over a quasi-circular merger.
Although the exact formation scenario in the studies on GW190521 
\cite{Gamba_2022,Gayathri:2020coq} are distinct, both 
agree that this event is the result of a dynamical formation scenario.

These recent studies \cite{RomeroShaw_2020,RomeroShaw_2021,Romero-Shaw:2022xko,romero-shaw_2025,
Gupte:2024jfe,Iglesias_2024,Gayathri:2020coq,Gamba_2022} have identified 
several GW events that are potentially produced by binaries with a 
non-negligible orbital eccentricity entering the LVK sensitivity band. Some 
of the candidates deserving further study include GW190620 
\cite{RomeroShaw_2021}, GW191109, GW200208\_22 \cite{Romero-Shaw:2022xko}, 
GW190701 and GW200129 \cite{Gupte:2024jfe}. 

This work aims to improve and build upon previous results by using model-based
parameter estimation with the eccentric version of \teo{} 
\cite{teogeneral_update} hand in hand with numerical-relativity simulations 
from the RIT catalog \cite{RIT_NR1,RIT_NR2,RIT_NR3,RIT_NR4} to probe high 
likelihood regions of parameter space. We study in particular GW200208\_22, 
first through parameter estimation using \teo{} with RIFT \cite{rift}. Then 
we generate new targeted numerical relativity simulations with the parameters 
of the highest likelihood models for use in parameter estimation via RIFT with 
the supplemented RIT catalog of simulations. This method of targeting high 
likelihood regions of parameter space to perform NR-based PE may potentially 
improve parameter estimates obtained through model-based methods. 
Additionally, we use the RIT catalog along with these targeted simulations to 
analyze GW190620 as a demonstration that our parameter space coverage allows 
for reasonable parameter estimates, even in the absence of specifically 
targeted simulations.

This paper is organized as follows. In \secref{s:methods}, we give an overview 
of our methods. In particular, in \secref{sec:FN} we present the numerical 
techniques used to produce our simulations, in \secref{ss:rift} we review the 
use of RIFT in our study, \secref{ss:NRsims} and \secref{ss:waveformmodel} 
provide the bank of simulations used and introduce the waveform model \teo{}, 
with focus on how eccentricity is defined, and \secref{ss:data} provides the 
settings used for analysis. In \secref{s:events}, we present the results of 
NR- and model-based parameter inference for GW200208\_22 and GW190620, as 
well as a numerical relativity accuracy test along with a mode-by-mode 
comparison of our highest likelihood simulation for GW200208\_22 and a \teo{} 
waveform of the same parameters. In \secref{s:conclusion}, we summarize our 
results and conclude with remarks on future work. \appendref{s:kde} provides 
details on how the posterior peak and associated 90\% \revthree{confidence interval} are 
calculated. Lastly, \appendref{s:NRsimsparam} provides the parameters of the 
targeted simulation produced for this work.



\section{Methods}
\label{s:methods}

\subsection{Full Numerical Techniques}\label{sec:FN}

In order to perform the full numerical 
simulations of binary black holes we use the LazEv code\cite{Zlochower:2005bj}
which employs 8th order spatial finite differences \cite{Lousto:2007rj}, 4th 
order Runge-Kutta time integration, and a reduced \cite{Zlochower:2012fk} 
Courant factor $(c=dt/dx=1/4)$.

For setting up numerical initial data for binary black holes,
we regularly adopt the puncture
approach~\cite{Brandt97b} along with the {\sc  TwoPunctures}
~\cite{Ansorg:2004ds} code.
In order to locate apparent horizons during numerical evolutions
we use the {\sc AHFinderDirect}~\cite{Thornburg2003:AH-finding} 
and compute horizon masses from its area $A_H$.
Furthermore, we measure the magnitude of the horizon spins 
$S_H$, using the ``isolated horizon'' algorithm \cite{Ashtekar:2004cn}
as implemented in Ref.~\cite{Campanelli:2006fy}.

We also use the {\sc Carpet}~\cite{Schnetter-etal-03b} mesh refinement driver
to pinpoint the evolution of the black holes across the numerical domain.
{\sc Carpet} provides a ``moving boxes'' style of mesh refinement, 
where refined grids of fixed size are arranged about the
coordinate centers of the holes. These grids are then moved 
following the trajectories of the holes during the numerical simulation.

The grid structure of our mesh refinements have a size of the largest
box for typical simulations of $\pm400M$.  The number of points between 0
and 400 on the coarsest grid is XXX in nXXX (i.e. n100 has 100
points).  So, the grid spacing on the coarsest level is 400/XXX.  The
resolution in the wavezone is $100M/$XXX (i.e. n100 has $M/1.00$, n120
has $M/1.2$ and n144 has $M/1.44$) and the rest of the levels are
adjusted globally. For comparable masses and non-spinning black holes, the 
grid around one of the black holes ($m_1$) is fixed at $\pm0.65M$ in size and 
is the 9th refinement level. Therefore the grid spacing at this highest 
refinement level is 400/XXX/$2^8$. When considering small mass ratio binaries, 
we progressively add internal grid refinement levels \cite{Lousto:2020tnb}. 
Here we set units such that $M=m_1^H+m_2^H$ is the addition of the horizon 
masses.

The extraction of gravitational radiation from the numerical
relativity simulations is performed using the formulas (22) and (23)
from \cite{Campanelli:1998jv} for the energy and linear momentum
radiated, respectively, and the formulas in \cite{Lousto:2007mh}
for angular momentum radiated, all in terms of the extracted Weyl scalar 
$\Psi_4$ at the observer location $R_{obs}=113M$. In order to extrapolate
the observer location to infinity, we use the perturbative formulas in 
Ref.~\cite{Nakano:2015pta}. This extraction observer location seems to
provide an acceptable accuracy for the low-medium eccentricity studies
carried out in this paper, see detailed study in \cite{Ficarra:2024nro}.

We perform simulations of eccentric binaries from a given initial separation 
corresponding to the \revthree{apoapsis} $r_p=a_r (1+e_r)$, by dropping the initial 
tangential linear momentum $P_t$ by a factor $1-f$ from the quasicircular one 
$P_c$ at this \revthree{apoapsis}, i.e. $P_t=P_c(1-f)$. The relationship of $e_r(f)$ has 
been computed to 3.5 post-Newtonian (PN) order in \cite{Ciarfella:2022hfy}. To 
the lowest Newtonian order this relationship takes the simple form
$e_0 = 2f-f^2$, and this was used to characterize eccentric
simulations in the 4th release of the RIT BBH catalog  \cite{Healy:2022wdn}. 
Here we provide an analytic mapping to include up to 1.5 PN terms,
\vspace{-10 pt}
\begin{eqnarray}
    e_{1.5PN}&&= e_0 \left(1+\left(4-e_0 (\eta-2)\right)\frac{M}{r_p}\right)+\\
    && 2\, s_{eff}\left(3 (1-e_0)-\frac{3+e_0^2}{\sqrt{1-e_0}} \right)\left(\frac{M}{r_p}\right)^{\frac{3}{2}},\nonumber
\end{eqnarray}
where
\begin{equation}
    s_{eff}=\frac{\left[\left(4+3 q \right) \mathit{\chi_1^z} +q \left(3+4 q \right) \mathit{\chi_2^z} \right]}{4 \left(1+q \right)^{2}},
\end{equation}
and $\eta=q/(1+q)^2$ is the symmetric mass ratio, with $q=m_2/m_1$.
Note that here $s_{eff}=S_{eff}/M^2$ is as derived from the PN Hamiltonian in 
\cite{Damour:2001tu} and as described in Eqs. (5)-(6), and (28)-(29) of 
\cite{Healy:2018swt}, not to be confused with $\chi_{eff}$ that is rather 
related to $S_0/M^2=(q\chi_2+\chi_1)/(1+q)$.

Since the NR simulations benefit from the total mass $M$ invariance, we would 
only know the physical scale after the matching to a specific signal is 
performed. We can then {\it a posteriori} evaluate the eccentricity at a 
standard value, such as 20 Hz at \revthree{apoapsis}, from its reference NR value  
$f_{\rm{ref}}^{\rm{NR}}$, see \figref{fig:freqhlnl} for details 
on the NR reference frequency.
In order to do so, we use the formulas for the eccentricity evolution at the lowest 
post-Newtonian order from \cite{Peters:1964zz}, Eq. (5.11), to obtain
\begin{equation}\label{eq:e20}
\begin{split}
    e(20Hz)= & e_{NR} \left(\frac{f_{ref}^{NR}}{20 {\rm Hz}}\right)^{19/18} \\
    \times\Biggl[1 +
    &  \frac{19}{9}e_{NR}\left( 1 - \left(\frac{f_{ref}^{NR}}{20 {\rm Hz}}\right)^{19/18} \right)
    \Biggr],
\end{split}    
\end{equation}
valid for small eccentricities and in the inspiral post-Newtonian regime of the 
binary.

We use the analytic expression of $e_{1.5PN}$ to remap the $e_0$ 
eccentricities given in the RIT BBH Catalog, then convert the $e_{1.5PN}$ 
eccentricity to its corresponding eccentricity at the \revthree{apoapsis} frequency of 20 Hz $e_{20}$ using 
\eqnref{eq:e20} for use in the parameter estimation corner plots. 
We note that this conversion to $e_{20}$ is done after both stages of 
RIFT are complete as we must use the rescaled mass for each simulation.
We also provide a direct comparison to the full $e_{3.5PN}$ eccentricity 
estimation for the new 42 targeted eccentric nonprecessing runs 
in the \appendref{s:NRsimsparam}.

\revthree{Finally, we also note here that while the TEOB models reference frequency is, $f_{\text{avg}} = (f_a+f_p)/2$, that relates to the \revthree{apoapsis} frequency $f_a$ used as reference for NR simulations by $f_{\text{avg}} = f_a\,(1 + e^2)/(1 - e)^2,$
and for the sake of completeness we also introduce the relationship with the mean orbital frequency,
$f_a = f_{mean}\,\sqrt{1-e}/(1+e)^{3/2} \simeq f_{mean}\,(1-2e)
\simeq f_{avg}\,(1-2e), e<<1$.}

\subsection{RIFT}
\label{ss:rift}

A merging compact binary can be completely characterized by its intrinsic and 
extrinsic parameters. The intrinsic parameters, $\lambda$, refer to the 
component masses, component spins, eccentricity, and matter quantities. The 
seven extrinsic parameters ($\theta$) describe the spacetime location and 
orientation of the system, including right ascension, declination, luminosity 
distance, coalescence time, inclination, orbital phase, and polarization. We 
will express masses in solar mass units and dimensionless nonprecessing spins 
in terms of Cartesian components aligned with the orbital angular momentum 
$\chi_{i,z}$.

RIFT \cite{rift, Wysocki_2019, Wofford_2022} is an iterative process 
consisting of two stages to estimate the intrinsic and extrinsic parameters 
of the binary source. It compares gravitational wave data $d$ to predicted 
gravitational wave signals $h(\pmb{\lambda},\pmb{\theta})$.
In the first stage, for each $\lambda_\alpha$ from some proposed ``grid'' 
$\alpha = 1,2,\ldots,N$ of candidate parameters, RIFT utilizes parallel 
computing to evaluate a marginal likelihood
\begin{equation}
    \Lmarg = \int \mathcal{L}(\pmb{\lambda},\theta)p(\theta)d\theta.
\end{equation}

from the likelihood $\mathcal{L}(\pmb{\lambda},\theta)$ of the gravitational 
wave signal in a multi-detector network, accounting for detector response. This 
stage, called integrate likelihood extrinsic (ILE), provides point estimates
for $\ln \Lmarg$ using either Gaussian Process (GP) regression or random 
forests to interpolate a full posterior distribution over the intrinsic 
parameters. See the RIFT papers
\cite{rift, Wysocki_2019, Wofford_2022} for more detailed discussion.

\subsubsection{Model Based RIFT}
\label{sss:modelRIFT}
In the second iterative stage for a model-based run, RIFT first approximates 
$\mathcal{L}(\lambda)$ based on the set of evaluations 
$\{(\lambda_\alpha, \mathcal{L}_\alpha)\}$. Then, using this approximation, 
RIFT generates a full posterior distribution over the intrinsic parameters
\begin{equation}
    p_{\rm{post}}(\pmb{\lambda}) = \frac{\Lmarg p(\pmb{\lambda})}{\int d\pmb{\lambda}\Lmarg p(\pmb{\lambda})} 
\end{equation}
where $p(\pmb{\lambda})$ is the prior on the intrinsic parameters.

The posterior is fairly sampled to generate a new grid using adaptive Monte
Carlo techniques. The evaluation points and weights in that integral are 
weighted posterior samples, which are fairly resampled to generate conventional 
independent, identically distributed ``posterior samples.” This second stage is 
called construct intrinsic posterior or CIP. For further details on RIFT’s
technical underpinnings and performance, see 
\cite{rift, Wysocki_2019, Wofford_2022, Wagner_2025}.

\subsubsection{NR Based RIFT}
\label{sss:NRRIFT}
The second phase of RIFT must be slightly different when using numerical 
relativity simulations as grid points. Each NR simulation corresponds to a 
particular combination of the intrinsic parameters $\pmb{\lambda}$ (mass ratio, 
spin components, and eccentricity) but can be scaled to any value of total 
mass. Therefore the second phase of RIFT instead must use a one dimensional fit 
to generate a refinement grid near the peak $\lnL$, adaptively exploring 
the mass parameter space for each combination of the other intrinsic parameters 
which are available as NR simulations. Exploration in the other intrinsic 
parameters is limited to fixed points determined by which simulations are 
available in the catalog. 

\subsection{Simulations}
\label{ss:NRsims}
The fourth release of the RIT catalog of BBH simulations \cite{RIT_NR4} 
consists of 1881 accurate simulations, 824 in eccentric orbits with 
$0 < e \leq 1$, including 709 nonprecessing and 115 precessing; and 611 
nonprecessing and 446 precessing quasicircular/inspiraling binary systems with 
mass ratios $q = m_2/m_1$ in the range $1/128 \leq q \leq 1$ 
and individual spins up to $\chi_i = S_i/m_i^2 = 0.95$. The catalog also 
provides initial parameters of the binary, trajectory information, peak 
radiation, and final remnant black hole properties. The catalog includes all 
waveform modes $\ell \leq 4$ of the Weyl scalar $\psi_4$ \CarlosO{and the strain $h$ up 
to $\ell=5$ for eccentric BBH} (both 
extrapolated to null-infinity) and both are corrected for the center of mass 
displacement during inspiral and after merger \cite{RIT_NR4}. 
 
In addition to the public RIT catalog we also use the first release of 
eccentric nonspinning BBH simulations covering up to 25 orbits 
\cite{Ficarra:2024nro}. This catalog includes 30 simulations with five 
eccentricities, $e \in [0,0.45]$, and six different mass ratios, 
$q \in [0.11,1.0]$. Both of these catalogs \cite{RIT_NR4,Ficarra:2024nro} 
report initial eccentricity in terms of the eccentricity parameter, $f$, which 
results in the Newtonian definition of eccentricity $e_0=2f-f^2$. This measure 
of eccentricity can be converted to 3.5PN order using the methods described in 
\cite{Ciarfella:2022hfy}, which allows for a better comparison to 
eccentricities reported by the model \teo{}. \revthree{We note that our bank of 
simulations does not have an even coverage of 
parameter space and contains regions with very few simulations which may affect the 
posteriors we generate.}


\subsection{Waveform Model}
\label{ss:waveformmodel}
\begin{figure}[h!tbp]
    \includegraphics[width=0.45\textwidth]{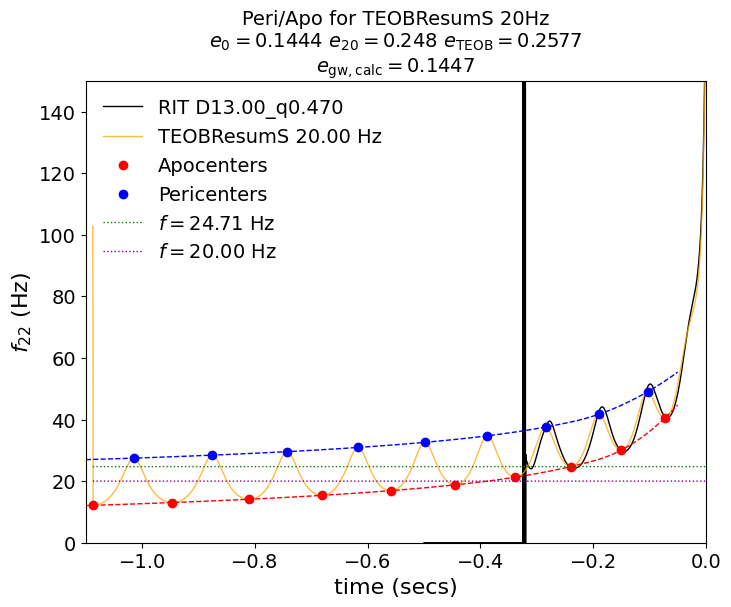}
    \caption{We ensure that a consistent definition of eccentricity is being
      used, particularly for follow-up numerical relativity simulations run
      specifically to target points near the peak likelihoods found via
      model-based parameter estimation for GW200208\_22. For the model-based 
      runs,
      the mean value of eccentricity at the reference frequency was used,
      whereas the RIT simulations use eccentricity define at \revthree{apoapsis} for a given
      frequency. This plot shows the values of eccentricity used by each given
      their particular definitions, as well as an overlap of the (2,2) mode to
      ensure that the shape of the waveforms match. 
      We note that for the simulation eGW::02, with total mass 
      $M=60.52 M_\odot$, the Newtonian eccentricity of $e_0=0.144$ translates 
      into an eccentricity at $20$ Hz of $e_{20}=0.248$.}
    \label{fig:gwecc_compare}
\end{figure}

\teo{} \cite{teobresums2} is an effective-one-body (EOB) model. This approach
combines the phases of two-body dynamics, including inspiral, merger, and
ringdown, into a single analytical method. First introduced by \cite{eob},
this method allows for highly accurate waveform calculations by mapping the
dynamics of a binary system onto one single, effective object that moves in the
potential. The effective object is described by equations of motion derived from
general relativity. \teo{} is informed by quasi-circular NR simulations 
of BBH coalescence events for calibration and for eccentric inspirals--validated by eccentric NR simulations. For the model-based parameter estimation
portions of this study, we use the eccentric spin-aligned waveforms from \teo{}
\cite{teogeneral_update} which were verified by NR for eccentricities up to
$e \leq 0.3$ at $10$ Hz. This model also incorporates higher order modes up
to $\ell = |m| = 5$, except for $m=0$. The user can also define the reference
frequency for \teo{} at \revthree{periapsis}, \revthree{apoapsis}, or the mean of both. 
The model-based parameter estimation runs using \teo{} for this study used the mean 
frequency as the reference frequency for eccentricity, whereas the NR 
simulations used the \revthree{apoapsis} frequency.
To provide a common benchmark to directly compare the different definitions of 
eccentricity used in TEOB and NR, we use the \textsc{gw\_eccentricity} package  
\cite{Shaikh_2023} which extracts an estimate of eccentricity from the
frequency modulations of the dominant (2,2) mode.  \figref{fig:gwecc_compare} 
shows a head-to-head comparison of the frequency versus time for the (2,2) mode 
extracted from an NR simulation and a corresponding TEOB estimate which most 
resembles it; this figure's title provides the corresponding eccentricities 
identified by these four methods: direct definition at the initial 
\revthree{apoapsis} frequency of the NR simulation $e_0$, \revthree{3.5 PN eccentricity at an apoapsis frequency of 20 Hz $e_{20}$}, direct definition at a mean 
frequency of $20$ Hz for \teo{} $e_{\rm{TEOB}}$, and the 
\textsc{gw\_eccentricity} estimate $e_{\rm{gw,calc}}$ \revthree{calculated from 
the \teo{} model at the NR \revthree{apoapsis} reference frequency}. Due to convention 
differences, the intrinsic TEOB parameterization adopts a higher value of its 
eccentricity parameter to explain this signal than the \textsc{gw\_eccentricity} 
or NR simulation conventions \revthree{for the initial $e_0$}, which in this case closely agree. 

The version of \teo{} used in this study to provide a benchmark analysis is 
a previous version of the model which includes only two parameters to determine 
the initial conditions, $e_0$ and $f_{\rm{ref}}$, to evolve the trajectories of 
the binary components. In the time since this analysis was performed, a new 
version of \teo{} \cite{nagar_2024_teo_dali,Nagar:2024oyk} has been released which 
includes the mean anomaly as a waveform parameter. Note that the value of the 
waveform eccentricity is different than the physical, time-varying eccentricity 
of the system, and may also mean different things depending on the choice of 
model. We specifically employ the eccentric branch of \teo{} \cite{teobresums2} which can be found here \url{https://bitbucket.org/teobresums/teobresums/src/1b965da2daf470d0a3227c1573a382b2eb7e7feb/?at=attic%2Feccentric}.

\subsection{Analysis} 
\label{ss:data}
We analyze publicly available data from GWTC-2.1, and GWTC-3 
\cite{LIGO-O3-O3a_final-catalog,LIGO-O3-O3b-catalog} and use the noise power 
spectral density curves (PSDs) associated with each specific event as part of 
the public data release.  For GW200208\_22 we used a low-frequency cutoff of 20 
Hz, and for GW190620 a 10 Hz cutoff. The high frequency cutoff was 448 Hz for 
both GW200208\_22, and GW190620. For details on the choice of high-frequency 
cutoff see Appendix E of \cite{LIGO-O3-O3b-catalog}, and Section V of 
\cite{LIGO-O3-O3a_final-catalog}, respectively for each event. We adopt a 
uniform prior for eccentricity over the range $e \in [0.0,0.9]$ for all events. 
We assume both spins are aligned with the orbital angular momentum and 
uniformly distributed in our prior range, which is either $|\chi_{i,z}|<0.5$ or 
$<0.9$, depending on the analysis used. The more restricted spin prior is used 
to constrain our posteriors under the alternative assumption that the events 
analyzed are unlikely to be highly spinning. All other extrinsic and
intrinsic priors are conventional: for example, we use a uniform prior in the 
detector frame masses $m_{i}$, modulo upper and lower limits in $\mc$ and
$q$; a Euclidean distance prior on $d_L$; and conventional uninformed priors on 
all extrinsic angles and event time. We utilize the nonprecessing simulations, 
both eccentric and non-eccentric, with symmetric mass ratio 
$\eta \in [0.045,0.25]$ (which corresponds to $q\in[0.05,1.0]$) 
rescaled over the detector frame chirp mass range of $\mc \in [20,40] M_\odot$ 
from the RIT catalog \cite{RIT_NR4} as an initial grid for NR-based PE. 

For the GW200208\_22 event, improvement of initial NR-based PE requires 
targeted NR simulations, these simulations have the intrinsic parameters of the 
highest $\lnLmarg$ points of model-based PE with spins in the range 
$\chi_{i,z} = \pm 0.4$. These targeted simulations are constructed using the 
same techniques as \cite{Ficarra:2024nro}, which differ slightly from the 
techniques described in \secref{sec:FN}. To further fill in the NR simulation 
grid in $e_0$ and $\chieff$ space, we add simulations of fixed $q=0.5$, 
$\chi_{2,i}=0$ at two values of eccentricity, $e_0=0.1,0.19$, with seven 
values of $\chi_{1,z} \in [0.1-0.6]$. In addition, two more simulations are 
added to fill a void in the grid in $q$ and $e_0$ space, with positive spin in
$m_1$, eccentricity of 0.049, and mass ratios of 0.6, and 0.7. 
Finally, we add 3 simulations to fill $(q,\chieff)$ space, where $q=0.85$ and $\chieff =\pm0.37,0$.

The parameters of these additional simulations are provided in 
\Tabref{tab:eGWidparameters} and \Tabref{tab:eGWfinalproperties} with the 
targeted simulations labeled eGW::01-23, additional simulations in 
$(e_0,\chieff)$ space labeled eGW::24-37, additional simulations in 
$(q,e_0)$ space labeled eGW::38-39, and simulations filling 
in $(q,\chieff)$ space are labeled eGW::40-42. The catalog of 30 eccentric nonspinning 
simulations covering up to 25 orbits \cite{Ficarra:2024nro} is employed in the 
resolution study in \secref{ss:nracc}, as well as to further fill in the grid 
of simulations for NR-based PE in \secref{s:events}. 

\section{Analysis of actual events}
\label{s:events}
In this section we present the results of the eccentric reanalysis of 
GW200208\_22 and GW190620 using direct comparison to numerical relativity. 
These events were analyzed based on prior evidence or indications of 
eccentricity \cite{Romero-Shaw:2022xko,RomeroShaw_2021,Gupte:2024jfe}.
While in the discussion below we employ comparisons to numerical 
relativity simulations, for context we also provide a comprehensive comparison with model-based 
results in \secref{ss:200208_model_NR}.

\subsection{GW200208\_22}
\label{ss:200208PE}


The LVK collaboration identified in GWTC-3 \cite{LIGO-O3-O3b-catalog} the low 
SNR event ($7.4_{-1.2}^{+1.4}$) GW200208\_22 as an event with high-mass 
sources, where the component masses, mass ratio and chirp mass in the detector 
frame as reported in \cite{O3b_PE_data_release} are: 
$m_1=83.4_{-48.7}^{+171.8}$, $m_2=21.9_{-11.8}^{+13.0}$, 
$q=0.21_{-0.16}^{+0.67}$, and $\mc=31.3_{-8.5}^{+26.2}$. This event also has 
significant support for positive effective spin with a 95\% probability that 
$\chieff>0$ where the reported effective spin from Table IV of 
\cite{LIGO-O3-O3b-catalog} is $\chieff=0.45_{-0.44}^{+0.43}$

\begin{figure*}[h!tbp]
    \includegraphics[width=1.5\columnwidth]{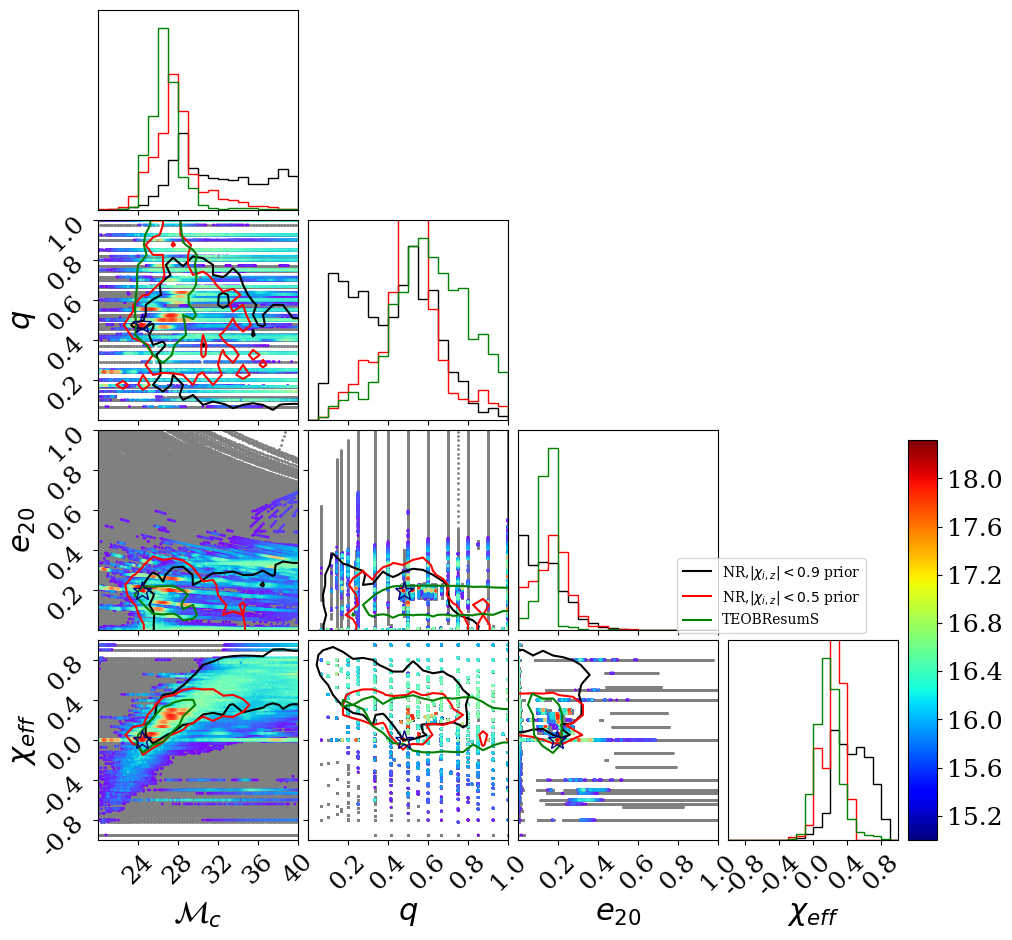}
    \caption{A corner plot showing the results of NR based PE for 
    GW200208\_22 with a grid consisting of the RIT catalog \cite{RIT_NR4}, 
    targeted NR simulations (parameters provided in \appendref{s:NRsimsparam}), 
    and eccentric non-spinning simulations \cite{Ficarra:2024nro} 
    for a total of 1248 simulations rescaled over the detector 
    frame chirp mass range of $\mc \in [20,40] \ M_\odot$ consistent with the 
    model-based 90\% \revthree{credible interval}. Two NR posteriors are shown with different 
    $\chieff$ priors, $|\chi_{i,z}|< 0.9$ (black) and $|\chi_{i,z}|< 0.5$ (red) with the 
    model-based posterior shown in green. Note that the TEOBResumS posterior 
    has eccentricity defined as $e_{\rm{TEOB}}$ at a mean frequency of 20 Hz 
    \revthree{but we convert $e_{\rm{TEOB}}$ to an apoapsis frequency of 20 Hz using the same methods as \figref{fig:gwecc_compare} and denote the converted value as $e_{\rm{TEOB,apo}}$.}
    The colorbar represents the value of $\lnLmarg$ and a star is placed at the peak $\lnL$ 
    simulation, these are present on all corner plots.}
    \label{fig:NRPE2}
\end{figure*}

\figref{fig:NRPE2} shows our analysis of GW200208\_22 using direct comparison 
to nonprecessing numerical relativity simulations, as described in 
\secref{ss:data}, along with the model-based posterior. We perform 
this analysis on a grid covering a detector frame chirp mass range of 
$\mc \in [20,40] \ M_\odot$ which is consistent with the model-based 90\% 
\revthree{credible interval.} While our underlying catalog of nonprecessing eccentric 
NR simulations has only a few simulations which have optimal $e$ and 
$\chi_{\rm eff}$, our catalog and thus analysis specifically includes 
followup simulations targeted at the expected binary parameters. Looking 
first at the marginal likelihoods derived by comparing each NR simulation 
against the data, the best-fitting numerical relativity simulations for this 
event are all somewhat eccentric, and many indicate modest aligned spin. 
Conversely, the most eccentric simulations in our catalog consistently fit 
poorly, suggesting the eccentricity is very likely below $e_0\simeq 0.3$.  
Similarly, none of the best fitting simulations have negative $\chi_{\rm eff}$, 
suggesting a preference for positive spin, and most of the best fitting 
simulations have modest mass ratio $q>0.4$. 

Interpolating these sparsely sampled marginal likelihoods versus mass, aligned 
spins, and eccentricity, we infer a joint posterior distribution for these 
parameters. Keeping in mind potential systematic errors introduced by our 
coverage, we would estimate that the binary mass ratio posterior favors a 
binary with moderately asymmetric components, where the maximum of a kernel 
density estimate (KDE) on the 1-dimensional posteriors have reasonable 
agreement for results derived using the priors $|\chi_{i,z}|< 0.5$ and 
$|\chi_{i,z}|< 0.9$, $q=0.527_{-0.269}^{+0.264}$ and 
$q=0.525_{-0.410}^{+0.224}$ respectively, 
where the maximum KDE value is reported with the symmetric 90\% \revthree{confidence
interval} on the 1D posterior; the details of constructing the maximum KDE 
value and its errors are further described in \appendref{s:kde}. 
Similarly, the marginal eccentricity posterior seems to favor modest 
eccentricity, $e_{20} = 0.198_{-0.180}^{+0.119}$
for $|\chi_{i,z}|< 0.5$ and $e_{20} = 0.027_{-0.020}^{+0.296}$
for $|\chi_{i,z}|< 0.9$ and bound it above ($e_{20}\lesssim 0.3)$, for both 
spin priors. The $\chieff$ posterior includes zero effective spin with a 
slight preference for positive spin for both choices of prior peaking at 
$\chieff=0.283_{-0.275}^{+0.130}$ and $\chieff=0.291_{-0.145}^{+0.491}$
for the priors $|\chi_{i,z}|< 0.5$ and $|\chi_{i,z}|< 0.9$ respectively. 
For both choices of spin prior, the chirp mass posterior has a maximum at 
$\mc =27.568_{-3.313}^{+5.889} M_{\odot}$ for $|\chi_{i,z}|< 0.5$ and 
$\mc =28.408_{-2.153}^{+10.783} M_{\odot}$ for $|\chi_{i,z}|< 0.9$. 
We obtain similar parameter values for mass ratio 
$q$, detector frame chirp mass $\mc$, and effective spin $\chieff$ as the 
LVK reported values \cite{LIGO-O3-O3b-catalog,O3b_PE_data_release}. 
\begin{figure*}[h!tbp]
    \centering
    \includegraphics[width=1.66\columnwidth]{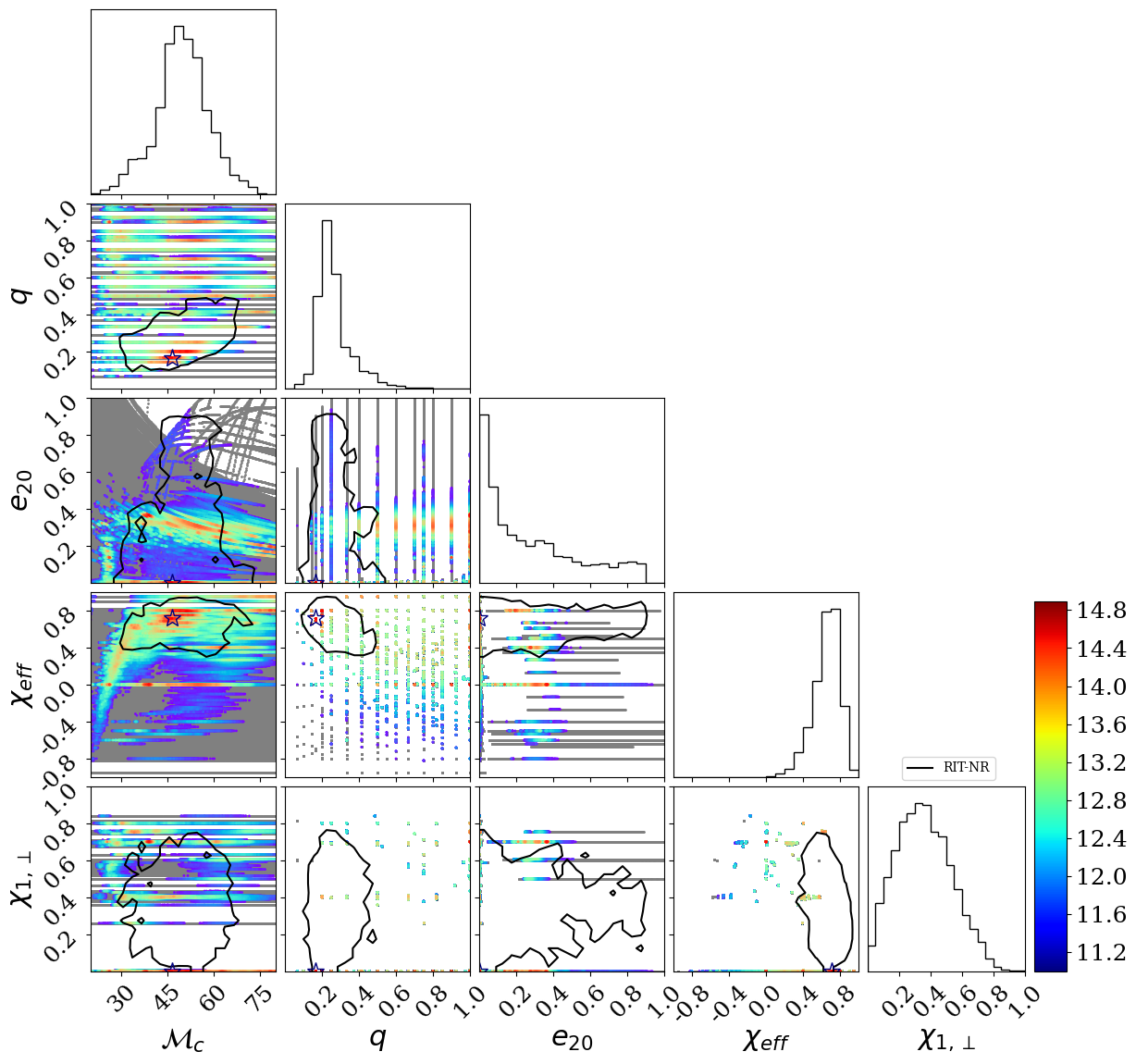}
    \caption{A corner plot showing the results of NR based PE for GW200208\_22 including precession parameter $\chi_{1, \perp}$ with a grid consisting of 1728 simulations from the RIT catalog \cite{RIT_NR4} with a detector frame chirp mass range of $\mc \in [20,80] \ M_\odot$. The posterior fit is generated using the priors described in \secref{ss:data} with an additional triangular prior on the precession parameter $\chi_{1, \perp}$. } 
    \label{fig:NRprecchi1_perp}
\end{figure*}
We note that the posterior with the less restrictive spin prior 
($|\chi_{i,z}|< 0.9$) is not well constrained over our chirp mass range due to a 
lack of simulations in the large $\chieff$, small mass ratio ($q<0.4$) region of 
parameter space. To account for this we move forward with the spin prior of 
$|\chi_{i,z}|< 0.5$ as our preferred posterior, since it also compares closer to that of the \teo{} model.

In addition to the NR-based analysis, \figref{fig:NRPE2} also shows the 
posterior found through model-based parameter estimation which was performed using 
\teo{} with the same priors as \secref{ss:data}, except for the 
spin prior which is $|\chi_{i,z}|< 0.99$. It is clear that the posteriors generated 
from models and the NR with a $|\chi_{i,z}|< 0.5$ spin prior agree quite well, 
as both methods favor a modest eccentricity, with the posterior of model-based 
PE peaking at \revthree{$e_{\rm{TEOB,apo}}= 0.149_{-0.082}^{+0.042}$} 
and bound it above at $e_{\rm{TEOB,apo}}\lesssim 0.3$.
\revthree{We note that $e_{\rm{TEOB,apo}}$ is $e_{\rm{TEOB}}$ converted to an apoapsis frequency of 20 Hz using the \textsc{gw\_eccentricity} package with the similar methods as \figref{fig:gwecc_compare} to compute eccentricity at an apoapsis frequency 20 Hz.} 
The $\chieff$ posteriors have almost 
identical maxima, as both methods include zero effective spin 
in their posterior with a slight preference for positive effective spin 
with the model-based posterior peaking at $\chieff = 0.191_{-0.224}^{+0.234}$. 
Similarly, the chirp mass posteriors for both model-based and NR-based PE 
return a similar maximum, with the model peaking at 
$\mc = 26.680_{-2.049}^{+4.300} M_{\odot}$. Finally, both posteriors 
favor a binary with moderately asymmetric components as there is strong support 
for $q > 0.4$ with the model-based PE maximum at $q=0.568_{-0.272}^{+0.354}$.
We provide a summary of the maximum posterior values found through a KDE analysis 
of the posteriors in \figref{fig:NRPE2} in \tabref{tab:200208_kde_params}. 

\figref{fig:NRprecchi1_perp} shows an analysis of GW200208\_22 using direct 
comparison to precessing, both eccentric and noneccentric, numerical relativity 
simulations, based on a comparison to the simulations provided in 
\cite{RIT_NR4}\footnote{The difference between the number of 1728 simulations used here and the 1881 of the catalog is due to the ranges in \revthree{$q \in [0.05,1.0], |\chi_{i,z}|<0.9, e \in [0.0,0.9]$}}
scaled over the detector frame chirp mass range of 
$\mc \in [20,80] \ M_\odot$. 
This analysis was completed with the same settings as the nonprecessing 
analysis but with additional triangular priors on the remaining spin components 
$|\chi_{i,x}|< 0.99$ and $|\chi_{i,y}|< 0.99$, as well as a triangular prior 
for the in-plane spin components, $\chi_{i, \perp} \in [0,1]$. 
Where the in-plane spin components are defined as
\begin{equation}
    \chi_{i, \perp} = \left|\frac{\mathbf{S_i}}{m_i^2} \times \mathbf{\hat{L}}\right|.
\end{equation}

\begin{figure*}[h!tbp]
    \includegraphics[width=1.5\columnwidth]{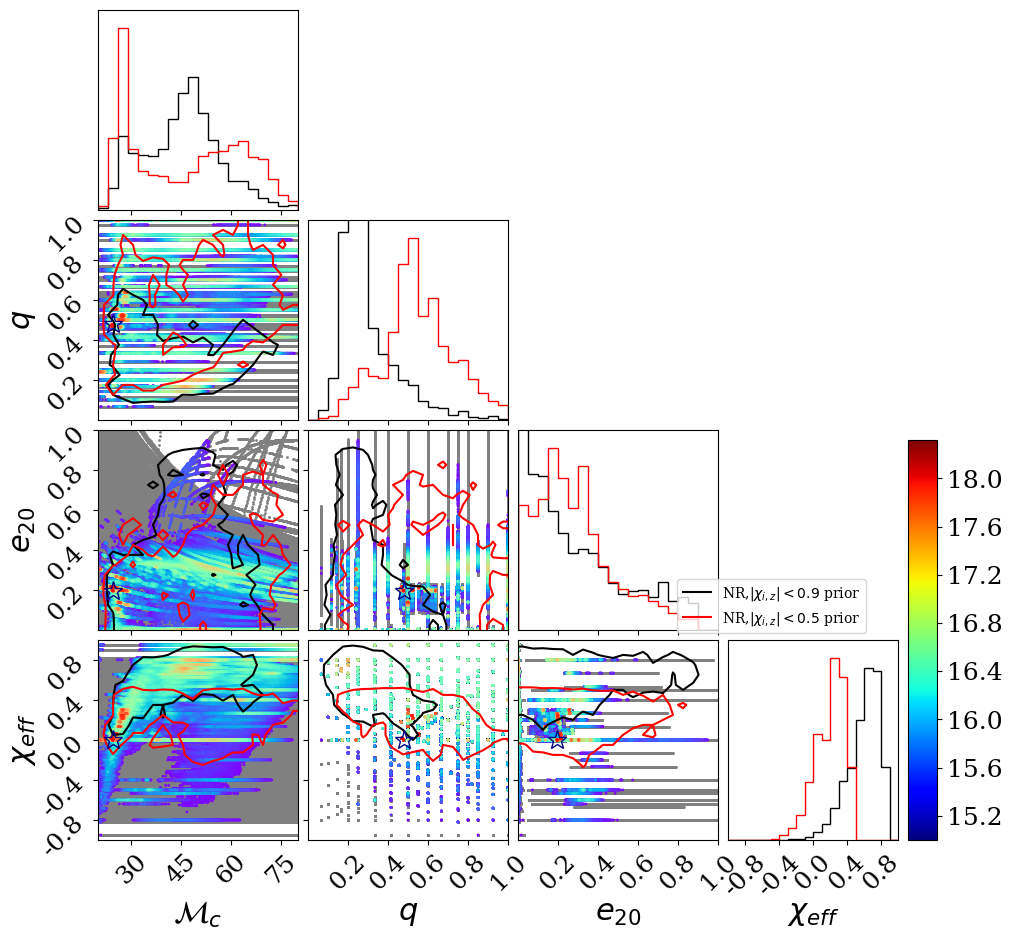}
    \caption{ A corner plot showing the results of NR based PE for 
    GW200208\_22 with a grid consisting of the RIT catalog \cite{RIT_NR4}, 
    targeted NR simulations (parameters provided in \appendref{s:NRsimsparam}), 
    and eccentric non-spinning simulations \cite{Ficarra:2024nro} for a total of 1248 simulations rescaled over the detector 
    frame chirp mass range of $\mc \in [20,80] \ M_\odot$. Two NR posteriors are shown with different 
    $\chieff$ priors, $|\chi_{i,z}|< 0.9$ (black) and $|\chi_{i,z}|< 0.5$ (red).}
    \label{fig:NR_mc20-80}
\end{figure*}

The marginal likelihoods obtained through direct comparison of each NR 
simulation to the data in this secondary precessing analysis, reveal the 
best-fitting numerical relativity simulations are non-eccentric with large 
positive effective spin.
Still some simulations with marginal likelihoods just below the best-fitting 
simulations have nonzero in-plane spins
and/or an anti-aligned spin for $m_2$ (i.e. negative), 
indicating spin-induced precession may be present in the binary. 
Contrary to the nonprecessing analysis the 
majority of best fitting simulations have mass ratio $q<0.4$, but
many have positive $\chieff$ which is consistent with the nonprecessing analysis, 
suggesting a preference for positive spin. 

Now we look at the joint posteriors generated by interpolating the marginal 
likelihoods versus mass, aligned spins, eccentricity, and the precession 
parameter $\chi_{1, \perp}$. The chirp mass and mass ratio posteriors do 
not agree with the nonprecessing analysis with a 
maximum near $\mc = 49.009_{-15.540}^{+14.948} M_{\odot}$ and
$q= 0.233_{-0.079}^{+0.237}$ respectively. The $\chieff$ posterior 
no longer includes zero effective spin with a preference for positive 
effective spin with a maximum at $\chieff = 0.696_{-0.312}^{+0.153}$.
The marginal eccentricity posterior is now bounded above at 
$e_{20}\lesssim 0.8$, but in the precessing analysis the maximum is 
at $e_{20}=0.052_{-0.046}^{+0.757}$. Seemingly this is due to the 
abundance of precessing simulations with $e_{20}=0$. 
Finally the $\chi_{1, \perp}$ posterior favors modest transverse spin, 
peaking at $\chi_{1, \perp} = 0.340_{-0.256}^{+0.329}$.
The large differences present in our parameter estimates compared 
to the aligned spin analysis, especially for the chirp mass and mass ratio are 
largely due to the difference in grid coverage over chirp mass, as the precessing 
analysis extends up to $\mc = 80 \ M_\odot$. \revthree{We find that this precessing 
analysis favors a simulation which is non-eccentric and highly spinning but not precessing 
as many of our highest $\lnL$ points have $\chi_{1, \perp}\approx 0$.} 
However the likelihoods returned by this analysis are significantly lower ($<15$) 
\revthree{due to the exclusion of the 42 targeted simulations and 30 simulations from 
\cite{Ficarra:2024nro}} and fall outside the 90\% CI found in the aligned analysis 
presented in \figref{fig:NRPE2}, suggesting that this event is better matched to an 
eccentric, \revthree{low spin} nonprecessing waveform.

\begin{figure*}[h!tbp] 
    \includegraphics[width=0.46\textwidth]{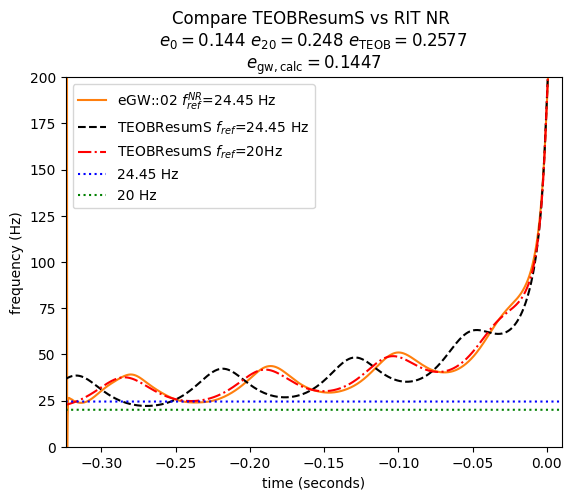}
    \includegraphics[width=0.485\textwidth]{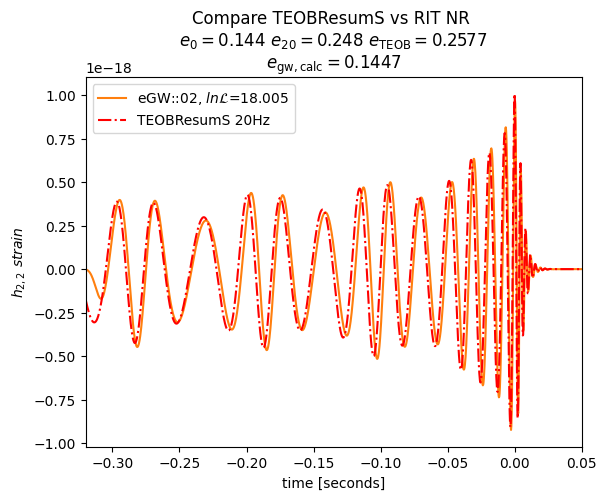}
    \caption{Comparison of frequency vs time (left), and $h_{2,2}$ strain vs 
    time (right) for highest $\lnL$ targeted simulation for GW200208\_22 
    (eGW::02 with total mass $M=60.52 M_\odot$) and \teo{} models with 
    the same intrinsic parameters. The left 
    panel demonstrates the difference in reference frequency definition, NR 
    defines the reference frequency at apoapsis while \teo{} models uses the 
    mean value of periapsis and apoapsis frequencies. The mean frequency of 20 
    Hz corresponds to the frequency $f_{\rm{ref}}^{\rm{NR}}=24.45$ Hz at apoapsis. 
    The right panel shows 
    the (2,2) mode GW strain of eGW::02 and the corresponding \teo{} 
    with $f_{ref}=20$Hz.
    }
    \label{fig:freqhlnl}
\end{figure*}

The aligned spin NR parameter estimation over the extended chirp mass range 
of $\mc \in [20,80] \ M_\odot$ shown in \figref{fig:NR_mc20-80} corroborates the 
parameters found in the precessing analysis of \figref{fig:NRprecchi1_perp} as 
the secondary peak in chirp mass near $\mc \approx 45 \ M_\odot$ is clearly present in 
the posterior generated with the spin prior $|\chi_{i,z}|< 0.9$ as well as in the log 
likelihood colormap in $(\chieff,\mc)$ parameter space. The majority of points 
making up this secondary mass peak have large positive effective spin and small mass 
ratio $q<0.4$, which is a region with sparse coverage by our simulation catalog, leading 
to a posterior which is not well constrained. Furthermore, these points fall 
below the peak likelihood region as found in both the model-based analysis and NR analysis,
as such we restrict our NR analysis to the highest likelihood region of chirp mass in the 
range $\mc \in [20,40] \ M_\odot$. With these things in mind, the preferred NR posterior 
for GW200208\_22 is the posterior generated from a grid over the mass 
range $\mc \in [20,40] \ M_\odot$ with the spin prior $|\chi_{i,z}|< 0.5$, this preferred 
posterior is shown in red in \figref{fig:NRPE2}.

\subsection{Discussion of GW200208\_22: differences between models and NR}
\label{ss:200208_model_NR}



\begin{figure*}[htbp]
    \includegraphics[width=1.5\columnwidth]{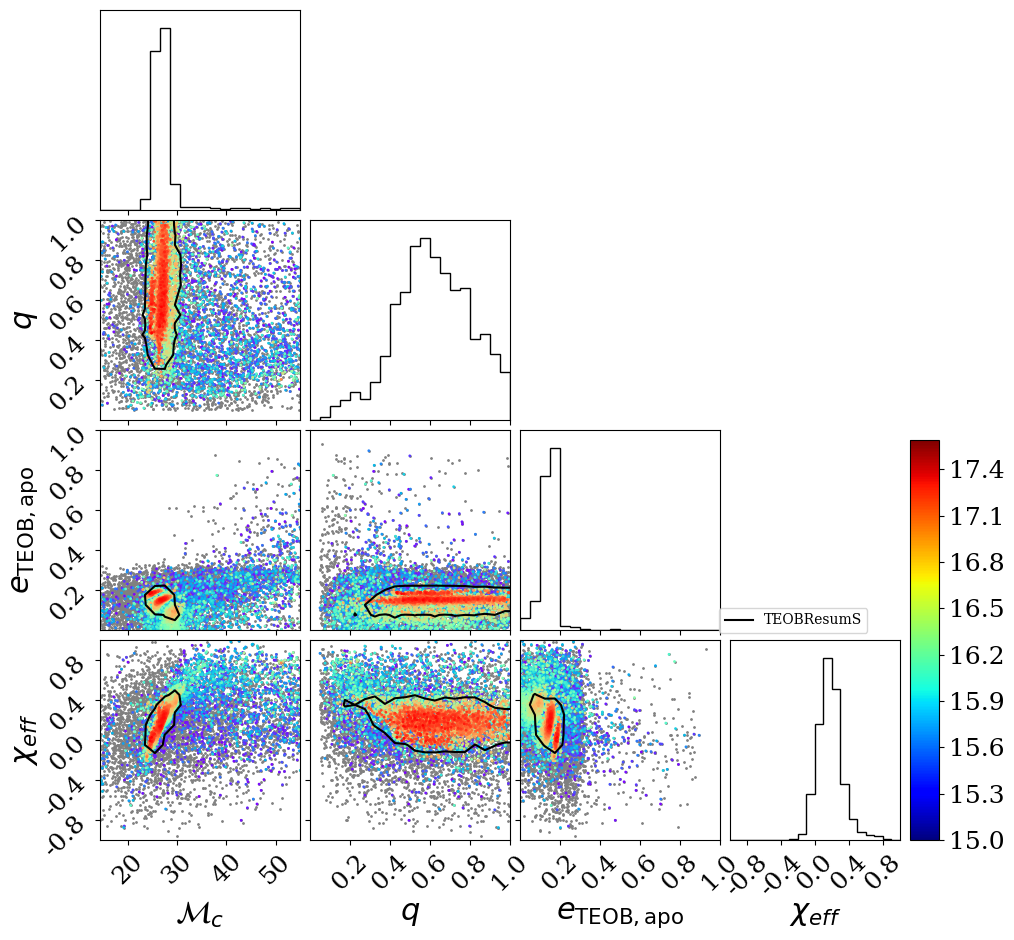}
    \caption{Model-based parameter estimation using RIFT with \teo{} for event GW200208\_22. \revthree{We note that $e_{\rm{TEOB,apo}}$ is $e_{\rm{TEOB}}$ converted to an apoapsis frequency of 20 Hz using the \textsc{gw\_eccentricity} package with the similar methods as \figref{fig:gwecc_compare} to compute eccentricity at an apoapsis frequency 20 Hz.}
    }
    \label{fig:modelPE1}
\end{figure*}

\figref{fig:freqhlnl} demonstrates the difference in how $f_{\rm{ref}}$ and 
eccentricity are defined for the numerical relativity simulations and the model 
\teo{}. The RIT NR simulations are constructed using a 
reference frequency defined at the apoapsis, this warrants that all waveform's 
higher frequencies will be included, while \teo{} constructs 
waveforms with the mean of the apoapsis and periapsis frequencies as its 
reference. 
Our \teo{} models used in analysis for GW200208\_22 have a 
mean reference frequency of 20 Hz, this happened to correspond to the apoapsis
frequency of 24.45 Hz for the simulation eGW::02 with total mass 
$M=60.52 M_\odot$, see \figref{fig:gwecc_compare}. Thus plotting a model with 
a reference frequency that numerically matches the NR reference frequency does 
not result in a matching frequency over time curve, as seen by mismatch of the 
model with a mean reference frequency of $f_{ref}=24.45 \rm{Hz}$ and the NR 
simulation in the left panel of \figref{fig:freqhlnl}.
We must ensure that the reference frequency used for generating models and NR
waveforms are equivalent by converting the mean frequency used in 
\teo{} to an apoapsis frequency for use in the NR simulations.
Taking the conversion into account results in a better match of frequency 
versus time for a model with a mean reference frequency of $f_{ref}=20 \rm{Hz}$ 
and the NR simulation.

In addition to the change in frequency plot, we look at the $(2,2)$-mode strain 
for a model with a 20 Hz mean reference frequency and the corresponding NR 
simulation in the right panel of \figref{fig:freqhlnl}. The model-based 
waveform and the NR waveform exhibit a good match during the inspiral phase and 
align nearly perfectly in the merger and ringdown phases. Further demonstrating 
that a reference frequency of 20 Hz defined as the mean of apoapsis and 
periapsis frequencies is equivalent to the 24.45 Hz frequency for this 
simulation at apoapsis, that the NR simulations use.

We show in \figref{fig:modelPE1} the model-based PE done with 
\teo{}, the posterior shown is the same as the red curve in 
\figref{fig:NR_mc20-80} along with the underlying marginal likelihoods derived 
by comparing \teo{} models against the data. This plot 
demonstrates the key difference when performing NR and model-based PE, 
models can be easily generated at any set of intrinsic parameters, while NR 
simulations are restricted to the set of intrinsic parameters at which they 
were generated, with the exception that they can scaled to any value of total 
mass. Additionally, the peak marginal likelihood points obtained from 
model-based PE are slightly lower than the marginal likelihoods of the same 
points obtained through NR-based PE. The slight improvement in marginal 
likelihood values may be due in part to the nature of how the NR 
simulations are constructed, as they are able to return more accurate waveforms 
for the higher order modes. 

\begin{figure}[h!tbp]
    \includegraphics[width=\columnwidth]{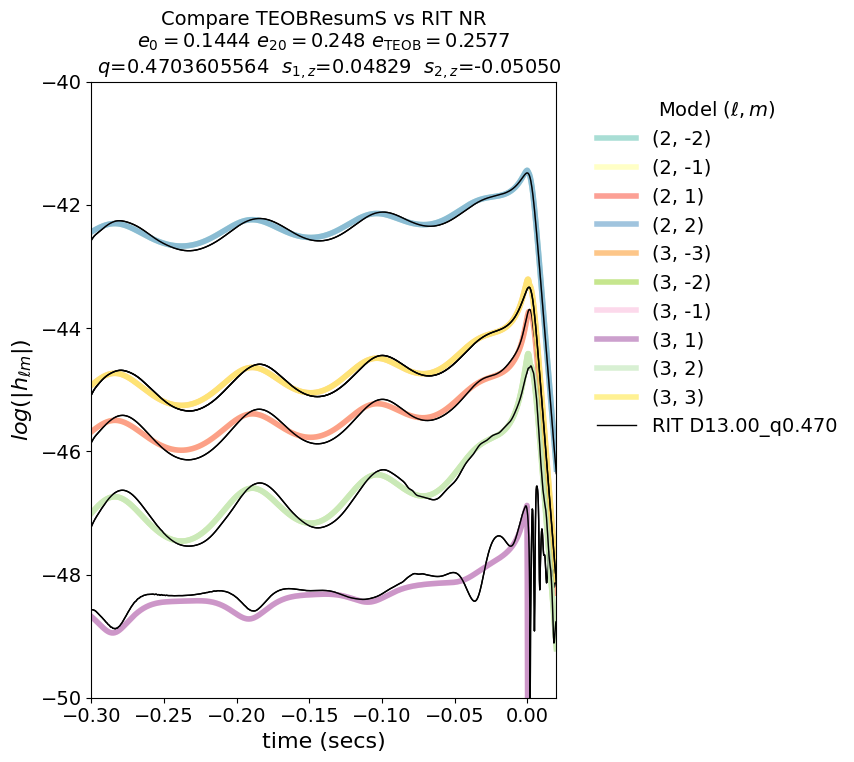}
    \caption{In addition to the (2,2) mode, we examine higher order modes to
      view a mode-by-mode comparison between the highest likelihood NR
      simulation (shown in black) with a \teo{} waveform generated with the
      highest likelihood parameter estimation results shown mode-by-mode in color. 
      We do this in particular to examine whether some
      differences in the waveform may cause the NR simulations to return
      slightly higher $\ln{\mathcal{L}}$ values than the models. We note that 
      for the simulation eGW::02 the Newtonian eccentricity of $e_0=0.144$ is 
      equivalent to the eccentricity at $20$ Hz of $e_{20}=0.248$. }
    \label{fig:modes}
\end{figure}

We examine this possibility with a mode by mode 
comparison of the peak targeted simulation (eGW::02). 
In order to visualize the potential differences between \teo{} modeling
and full numerical waveforms we display in Fig.~\ref{fig:modes} the comparison
of the leading $\ell=2,3$ modes with the top targeted simulation eGW::02. From 
this we observe that the key differences are not localized in any particular 
mode but rather in the generic higher accuracy of NR waveforms.
We also recall here the studies made in \cite{Lovelace:2016uwp} where it was 
found (See Fig.~3 there) that the completely different implementation of the 
full numerical solution to the two black hole problem, such as those of RIT and 
SXS, agree with each other \CarlosO{(up to the $\ell=5$ modes tested)} 
by two orders of magnitude closer than each of them with the SEOB model for the 
targeted runs of the first event GW150914.

\begin{table*} 
  \caption{Peak marginal likelihood, $\lnLmarg$, NR simulations and models of 
  GW200208\_22, where all have the dimensionless spin components 
  $s_1^x=s_1^y=s_2^x=s_2^y=0$ and $\lnL\geq 17$ as reported from the ILE stage 
  of RIFT. The likelihood values reported in the  NR\_PE $\lnL$ column are a 
  result of gaussian process fitting discussed in \secref{ss:200208_model_NR}.
  For the targeted simulations generated from the peak $\lnL$ points in model-
  based analysis, the model-based eccentricity $e_{\rm{TEOB,apo}}$ and 
  likelihoods are provided. For simulations not generated from the model-based 
  analysis (eBBH::08, eGW::26, eGW::36) these values are not provided 
  and denoted by -- in the $e_{\rm{TEOB,apo}}$ and model\_PE $\lnL$ columns.
  The NR starting frequency, $f_{\rm{ref}}^{\rm{NR}}$, is the frequency at 
  apoapsis which corresponds to the mean frequency of $f_{\rm{ref}}=20$ Hz 
  used by \teo{}, see \figref{fig:gwecc_compare}.
  $e_{20}$ is the conversion of the initial 3.5PN order eccentricity, 
  $e_{3.5\rm{PN}}$, to 20 Hz. Simulation eBBH::08 is EccBBH::08 from 
  \cite{Ficarra:2024nro}.
  }\label{tab:peakLnL} 
\begin{ruledtabular}
\begin{tabular}{cccccccccccccc}

NR\_sim\_id	&	$m_1/M_\odot$&	$m_2/M_\odot$&	$q=m_2/m_1$&	$s_1^z$&	$s_2^z$&	$e_{3.5\rm{PN}}$&  $e_{20}$& $e_{\rm{TEOB,apo}}$&	model\_PE&	NR\_PE& NR $f_{\rm{ref}}$\\
 & & & & & & & & &$\lnL$& $\lnL$& (Hz)\\
\hline
eBBH::08& 41.162& 19.659& 0.478& 0& 0& 0.241& 0.200& --&	--& 18.658& 15.22\\
eGW::02&  41.160& 19.360& 0.470& 0.0483& -0.0506& 0.216& 0.248& \revthree{0.181}& 17.338& 18.005& 24.45\\
eGW::04&  43.798& 22.778& 0.520& 0.2180& 0.2931& 0.203& 0.221& \revthree{0.151}& 17.153& 17.784& 22.48\\
eGW::08& 39.386& 21.786& 0.553& 0.2812& -0.3379& 0.212& 0.242& \revthree{0.187}& 17.087& 17.406& 24.33\\
eGW::06& 43.824& 22.953& 0.524& 0.3446& -0.0214& 0.204& 0.222& \revthree{0.155}& 17.392& 17.485& 22.54\\
eGW::07& 38.604& 21.366& 0.553& -0.0007& 0.1314& 0.213& 0.246& \revthree{0.188}& 17.009& 17.405& 24.69\\
eGW::26&  43.417& 21.708& 0.500& 0.2000& 0& 0.139& 0.152& --& --& 17.078& 22.34\\
eGW::36&  45.118& 22.560& 0.500& 0.4501& 0& 0.267& 0.288& --& --& 16.977& 22.58\\
\end{tabular}
\end{ruledtabular}
\end{table*}

We are able to directly compare these peak likelihoods due to the method we 
employed during the targeting stage of our analysis, whereby we generated 
simulations at the intrinsic parameters of the peak marginal likelihood points 
from the model-based PE. Exact intrinsic parameters and associated maximum 
marginal likelihoods found through a gaussian process fit for the simulations 
with $\lnL \geq 17$ in the ILE stage of RIFT are given in \Tabref{tab:peakLnL}, 
where likelihoods from both NR- and model-based PE are provided for only the 
targeted simulations due to the method we employed to generate targeted 
simulations. For the peak likelihood simulations generated through other means 
(without the input from model-based analysis) we do not generate 
models with equivalent intrinsic parameters for comparison with the NR-based 
analysis likelihoods.

\begin{figure}[h!tbp]
    \includegraphics[width=\columnwidth]{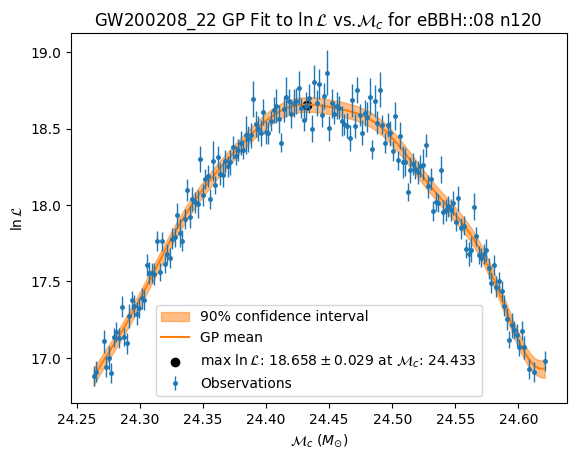}
    \caption{Gaussian Process fit for the top matching simulation (eBBH::08) at 
    resolution n120 for event GW200208\_22. RIFT data points and their errors 
    from the ILE stage are shown in blue. The mean prediction of the GP fit is 
    given by the orange curve, and its 90\% \revthree{confidence interval} shaded in 
    orange. The maximum $\lnL$ of the GP mean and its 90\% 
    \revthree{confidence interval} is given by the black point.
    }
    \label{fig:n120_gpfit}
\end{figure}

To eliminate possible data conditioning effects present due to the manner in 
which the NR simulations are loaded for use in RIFT, we perform a gaussian 
process fit on the simulations presented in \Tabref{tab:peakLnL} and 
\Tabref{tab:NRacc}. We rerun the ILE stage of RIFT for each simulation 
with the same settings as described in \secref{ss:data} but with the use of a 
denser initial grid in chirp mass with ten times the points, where 
we perform the fit in the chirp mass region which returned a maximal 
likelihood. We take the chirp mass limits to be $\pm 2$ of the chirp 
mass of each maximal likelihood simulation, for example eBBH::08 has 
$\mc=24.436 M_\odot$ thus the fit is performed on the region
$\mc =24.436M_\odot\pm 2 M_\odot$. The gaussian process fit is performed 
on the output from the first stage of RIFT (ILE), we first perform a cut 
in marginal log-likelihood, by only considering the data points that fall 
within 2 of the maximum $\lnL$, then we fit the marginal likelihoods as a 
function of chirp mass. We employ an additive kernel composed of a white 
noise kernel with variance $8\times10^{-5}$ plus the product of a constant 
kernel with value $0.5$ and the radial basis function kernel with length scale 
set at 2 times the standard deviation of chirp mass. We then compute the mean 
prediction of the GP fit and its 90\% \revthree{confidence interval}, then report the 
maximum $\lnL$ of the fit and its associated chirp mass. The results of this 
fitting method for eBBH::08 at resolution n120 are shown in 
\figref{fig:n120_gpfit}, where the maximum log-likelihood and its 90\% 
\revthree{confidence interval} of the GP fit is $18.658\pm0.029$ at a chirp mass of 
$24.433 \ M_\odot$.

\begin{table}[h]
\renewcommand{\arraystretch}{1.5}
  \caption{Summary of Maximum Kernel Density Estimates for the 
  posteriors of GW200208\_22 from \figref{fig:NRPE2}. Note that eccentricity 
  for \teo{} is $e_{\rm{TEOB,apo}}$.
  }\label{tab:200208_kde_params} 
\begin{ruledtabular}
\begin{tabular}{c|ccc}
  & Fig. 2 NR & Fig. 2 NR & Fig. 2   \\
Parameters & $|\chi_{i,z}|< 0.5$ & $|\chi_{i,z}|< 0.9$ & \teo{} \\  
  & posterior & posterior & \\
\hline
$\mc (M_\odot)$ & $27.568_{-3.313}^{+5.889}$ & $28.408_{-2.153}^{+10.783}$ & $26.680_{-2.049}^{+4.300}$ \\
$q$ & $0.527_{-0.269}^{+0.264}$ & $0.525_{-0.410}^{+0.224}$ &  $0.568_{-0.272}^{+0.354}$ \\
$e_{20}$ & $0.198_{-0.180}^{+0.119}$ & $0.027_{-0.020}^{+0.296}$ & \revthree{$0.149_{-0.082}^{+0.042}$} \\
$\chieff$ & $0.283_{-0.275}^{+0.130}$ & $0.291_{-0.145}^{+0.491}$ & $0.191_{-0.224}^{+0.234}$ \\
\end{tabular}
\end{ruledtabular}
\end{table}

\begin{figure*}[ht!bp] 
    \includegraphics[width=\textwidth]{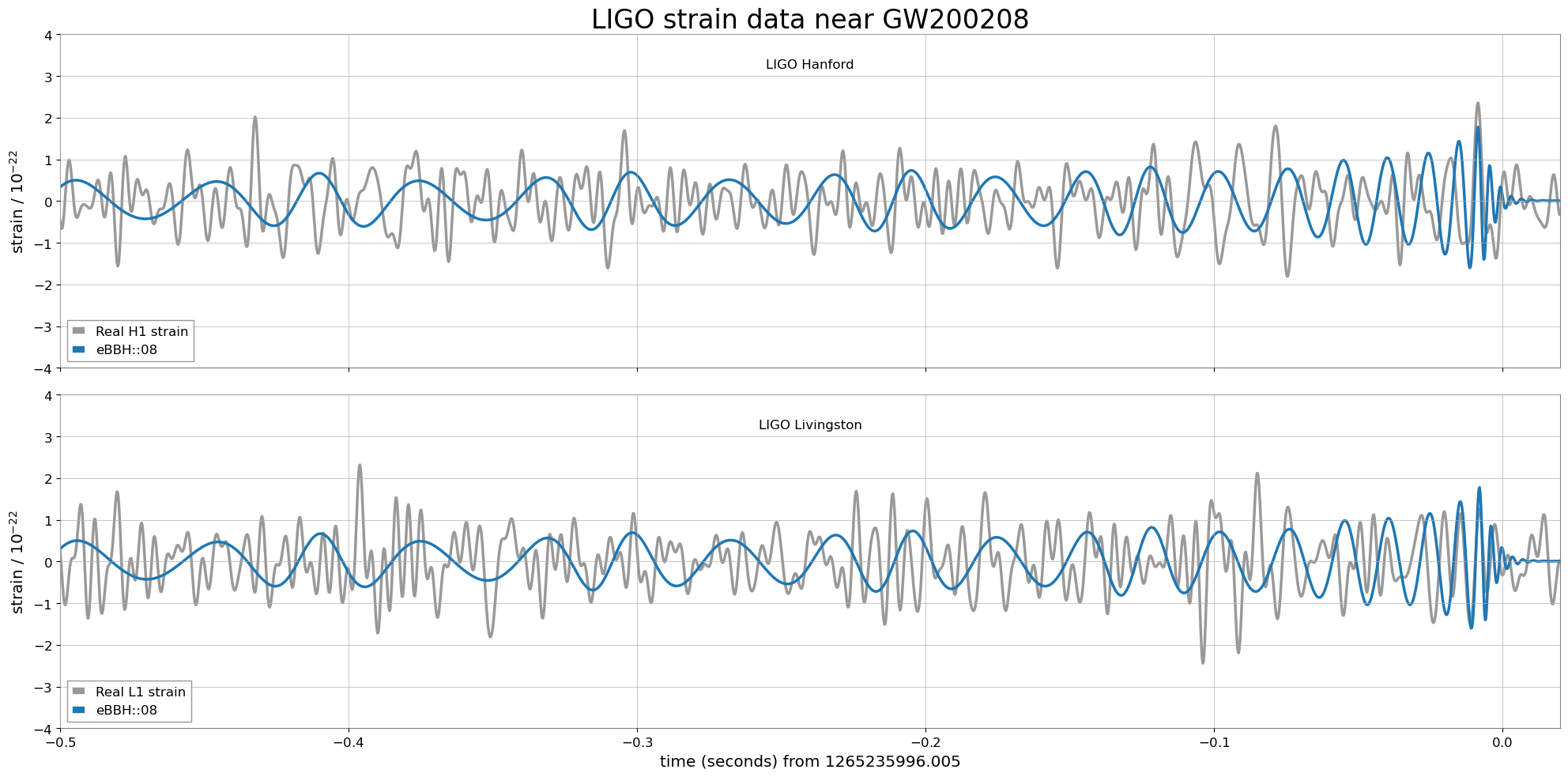}
    \caption{Best fitting time-domain NR waveform (eBBH::08) overlaid on whitened strain data near GW200208\_22.}
    \label{fig:200208waveform}
\end{figure*}




\subsection{Evaluation of Numerical Relativity accuracy}\label{ss:nracc}




The targeted NR simulations have been designed with the numerical grid 
structure and base n120 resolution optimized for catalog production as studied 
in \cite{Ficarra:2024nro}. In order to evaluate the accuracy of this reference 
numerical set up to match the gravitational wave signals detected by the LVK 
collaboration, we perform successive lower and higher global resolution 
(otherwise identical) simulations of eBBH::08, the top matching waveform of 
GW200208\_22. The results of evaluating $\lnL$ for these full numerical 
resolutions n100, n120, n144, n172 and n208, corresponding to global factors of 
1.2 increases, are displayed in Table~\ref{tab:NRacc}. We first note that the 
deviations of $\lnL$ from its n120 reference remain well within a fraction of a 
tenth of a percent and over an order of magnitude less than the differences we 
observe among the different targeted simulations in Table~\ref{tab:peakLnL}.
We also provide in Table~\ref{tab:NRacc} a direct comparison of the waveforms 
at different resolutions by computing their overlap over the whole range of the 
simulations as given by the matching measure,
\begin{eqnarray}
    \mathscr{M} \equiv \frac{\left<h_1\left|\right.h_2\right>}{\sqrt{\left<h_1\left|\right.h_1\right>\left<h_2\left|\right.h_2\right>}},
\end{eqnarray}
as implemented via a complex overlap as described in Eq.~(2) in 
Ref.~\cite{Cho:2012ed}:
\begin{equation}
    \left< h_1 \left|\right. h_2 \right>=2\,{\rm max_t}\left| \int_{-\infty}^{\infty} \frac{d\omega e^{i\omega t}}{S_n(\omega)}\left[\tilde{h}_1(\omega) \tilde{h}_2(\omega)^* \right]\right|,\label{eq:match}
\end{equation} 
where $\tilde{h}(\omega)$ is the Fourier transform of $h(t)$ and $S_n(\omega)$ 
is the power spectral density of the detector noise (here specifically taken 
$S_n(\omega)=1$ since we are interested in the direct waveforms comparisons).
We adopt the leading modes $(\ell,m)=(2,2)$ of $\psi_4$ for the computations 
and we do maximize over an overall constant time shift but not an overall 
constant phase shift.

\begin{table}
  \caption{$\lnL$ and matching for the peak likelihood NR simulation (eBBH::08) 
  of GW200208\_22. The likelihood values reported here are a result of gaussian 
  process fitting discussed in \secref{ss:200208_model_NR}. Our numerical resolution reference is n120.
  }\label{tab:NRacc} 
\begin{ruledtabular}
\begin{tabular}{ccccc}
Resolution & $\lnL$& $\Delta(\lnL)$ & matching \\
\hline

n100 & 18.659 & -0.010 &  0.971862\\
{\bf n120} & 18.658 & -0.011 &  0.999576 \\
n144 & 18.653 & -0.016 &  0.994080 \\
n172 & 18.676 & +0.007 & 0.999890 \\
n208 & 18.669 &\ 0.000 & 1.000000\\
\end{tabular}
\end{ruledtabular}
\end{table}


Figure~\ref{fig:200208waveform} displays the highest marginalized log-
likelihood NR waveform (top candidate eBBH::08 in Table~\ref{tab:peakLnL}) 
overlaid with whitened strain data near GW200208\_22 detector response in the 
LIGO Hanford (H1) and Livingston (L1) instruments.  
We note that the fit here displays nearly 10 waveform cycles before the binary 
black hole merger and thus contributes to the case of eccentric mergers with 
more of an inspiral orbit in the LIGO band compared to the other strong case 
for eccentric merger, GW190521, as this is much more massive 
\cite{Gayathri:2020coq} than GW200208\_22.

\subsection{GW190620}
\label{ss:190620}

The LVK collaboration identified in GWTC-2.1 \cite{LIGO-O3-O3a_final-catalog} 
GW190620 an event with SNR $12.1_{-0.4}^{+0.3}$ where the component 
masses, mass ratio and chirp mass in the detector frame as reported in 
\cite{O3a_PE_data_release} are: $m_1=84.6_{-15.4}^{+20.0}$, 
$m_2=53.1_{-19.7}^{+17.1}$, $q=0.62_{-0.27}^{+0.33}$, and 
$\mc=57.6_{-11.2}^{+9.0}$. This event also has significant support for positive 
effective spin where the reported effective spin from Table VI of 
\cite{LIGO-O3-O3a_final-catalog} is $\chieff=0.33_{-0.25}^{+0.22}$.

\begin{figure*}[h!tbp]
    \includegraphics[width=1.5\columnwidth]{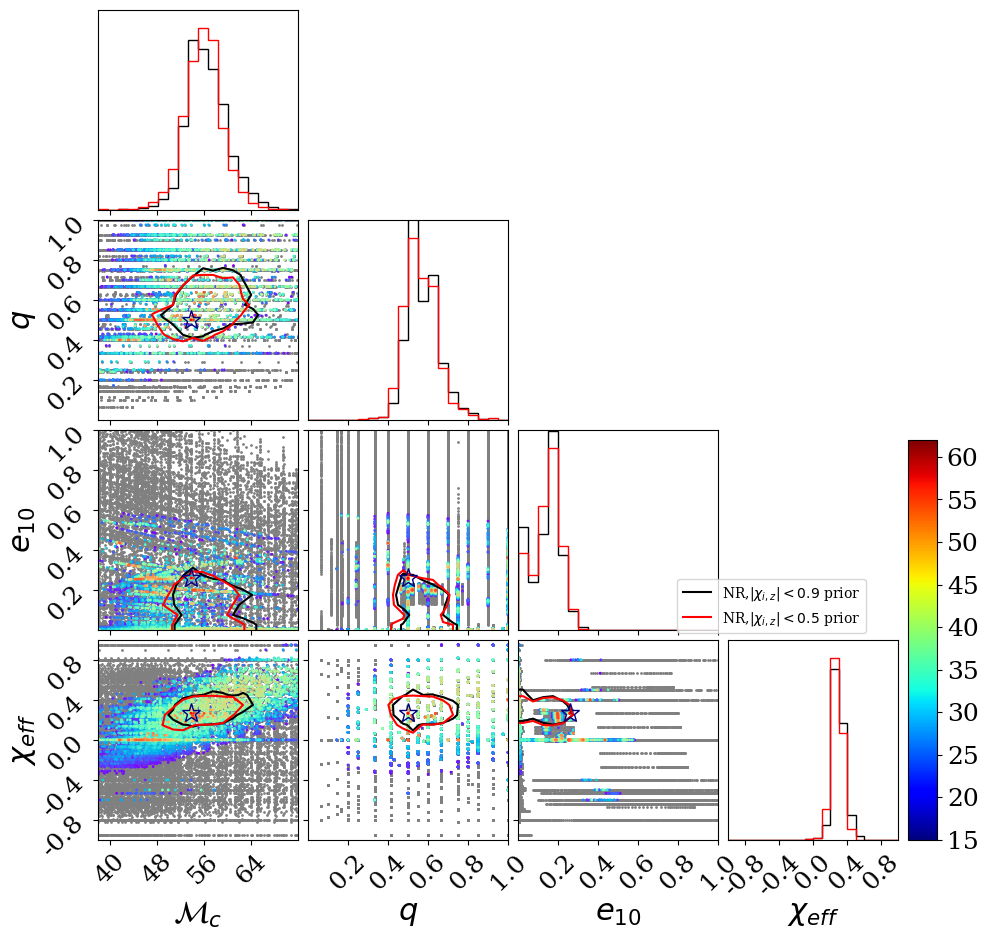}
    \caption{A corner plot showing the results of NR based PE for GW190620 
    using a grid consisting of the RIT catalog \cite{RIT_NR4}, 
    targeted NR simulations generated for 
    GW200208\_22 (parameters provided in \appendref{s:NRsimsparam}), 
    and eccentric non-spinning simulations \cite{Ficarra:2024nro} for a total of 1201 
    simulations. Two posteriors are shown with different spin priors, $|\chi_{i,z}|< 0.9$ 
    (black) and $|\chi_{i,z}|< 0.5$ (red). We provide eccentricity at the 
    reference frequency for this event of 10 Hz, which is found by using 
    \eqnref{eq:e20} with 10 Hz in the denominator in place of 20 Hz. }
    \label{fig:190620NRPE}
\end{figure*}

\begin{table*}[hbtp] 
  \caption{Peak marginal likelihood, $\lnLmarg$, NR simulations of 
  GW190620, where all have the dimensionless spin components 
  $s_1^x=s_1^y=s_2^x=s_2^y=0$ and $\lnL\geq 60$ as reported from the ILE stage 
  of RIFT. The likelihood values reported in the  NR\_PE $\lnL$ column are a 
  result of gaussian process fitting discussed in \secref{ss:200208_model_NR}.
  The NR starting frequency, $f_{\rm{ref}}^{\rm{NR}}$, is the frequency at 
  apoapsis which corresponds to the mean frequency of $f_{\rm{ref}}=10$ Hz 
  used by \teo{}, see \figref{fig:gwecc_compare}. 
  $e_{10}$ is the conversion of the initial 3.5PN order eccentricity, $e_{3.5PN}$, to 10 Hz at \revthree{apoapsis}. Likewise, $e_{20}$ is the conversion of the initial 3.5PN order eccentricity, $e_{3.5PN}$, to 20 Hz at \revthree{apoapsis}.
  }\label{tab:190620peakLnL} 
\begin{ruledtabular}
\begin{tabular}{cccccccccccccc}

NR\_sim\_id	&	$m_1/M_\odot$&	$m_2/M_\odot$&	$q=m_2/m_1$&	$s_1^z$&	$s_2^z$&	$e_{3.5\rm{PN}}$&  $e_{10}$& $e_{20}$& NR\_PE $\lnL$& NR $f_{\rm{ref}} (Hz)$\\
\hline
eGW::35& 88.421& 44.211& 0.500& 0.4001& 0& 0.269& 0.284& 0.188& 61.954& 11.53\\
eGW::06& 87.042& 45.590& 0.524& 0.3446& -0.0214& 0.204& 0.218& 0.133& 60.991& 11.25\\
eGW::20& 80.980& 51.651& 0.638& 0.3963& 0.0584& 0.201& 0.215& 0.131& 60.800& 11.25\\
eGW::04& 87.005& 45.249& 0.520& 0.2180& 0.2931& 0.203& 0.217& 0.132& 60.917& 11.28\\
RIT-eBBH-1828& 70.429& 70.391& 0.999& 0& 0.8009& 0.528& 0.381& 0.213& 60.543& 4.66\\
RIT-eBBH-1447& 77.281& 30.912& 0.400& 0& 0& 0.560& 0.477& 0.282& 60.501& 5.79\\
RIT-eBBH-1221& 73.333& 36.667& 0.500& 0& 0& {0.260}& 0.276& 0.230& 60.363& 16.06\\

\end{tabular}
\end{ruledtabular}
\end{table*}

\begin{figure*}[htb!p] 
    \includegraphics[width=\textwidth]{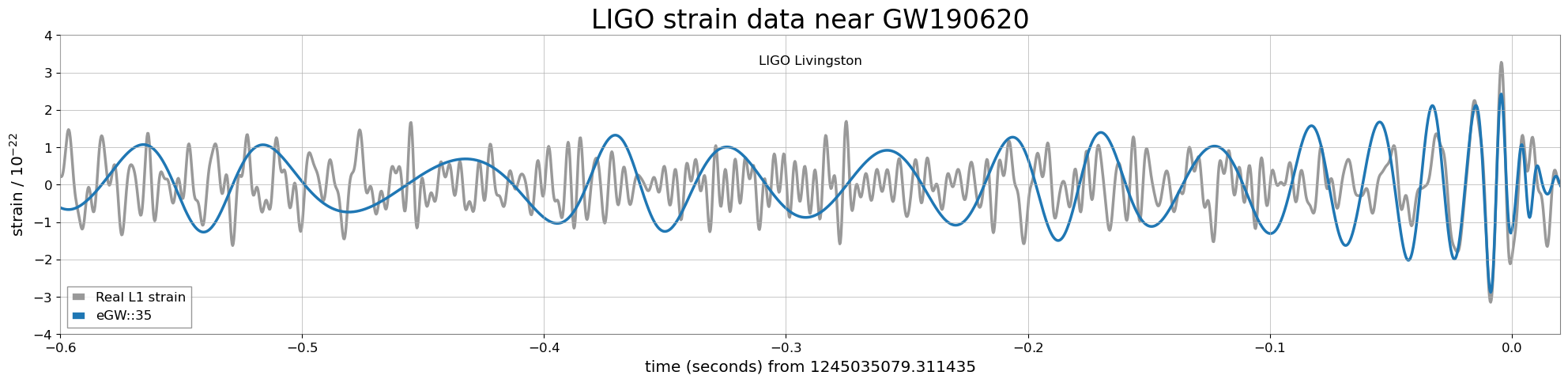}
    \caption{Best fitting time-domain NR waveform (eGW::35) overlaid on whitened LIGO Livingston (L1) strain data near GW190620.}
    \label{fig:190620waveform}
\end{figure*}

\figref{fig:190620NRPE} shows the results of parameter inference for GW190620, 
using direct comparison to the bank of nonprecessing numerical relativity 
simulations described in \secref{ss:data}. \revthree{Note that we use only the 
GW190620 L1 data from the LIGO Livingston detector, as H1 of LIGO Hanford 
was not observing and we exclude Virgo data as it is not sensitive enough.} 
Unlike in the case of GW200208\_22, 
numerical relativity simulations specifically targeted for this event were not 
generated as part of the analysis for GW190620. This was done to demonstrate 
that the bank of simulations described in \secref{ss:data} provides 
sufficient coverage of parameter space to return reasonable parameter 
estimates. 

We first look at the marginal likelihoods computed from the comparison of each 
NR simulation against the data. For this event all of the best-fitting 
numerical relativity simulations are somewhat eccentric, with many having 
moderate aligned spin. Again, the highly eccentric simulations in our catalog 
consistently fit the data poorly, indicating that eccentricity is likely 
bounded below $e_{10}\lesssim0.3$. Likewise, most of the best fitting 
simulations exhibit a modest mass ratio $q > 0.4$, and none of the best fitting 
simulations have negative $\chieff$, implying a preference for positive spin. 
Exact intrinsic parameters and associated maximum 
marginal likelihoods found through a gaussian process fit (see 
\figref{fig:n120_gpfit} discussion) for the simulations 
with $\lnL \geq 60$ in the ILE stage of RIFT are given in \Tabref{tab:190620peakLnL}.

These sparsely sampled marginal likelihoods are interpolated over mass, aligned 
spins, and eccentricity resulting in an inferred joint posterior distribution 
for these parameters. Noting that potential systematic errors may have been 
introduced by our coverage, we estimate that the binary mass ratio posterior 
favors a binary with moderately asymmetric masses, peaking at 
$q=0.524_{-0.075}^{+0.182}$ and $q=0.523_{-0.056}^{+0.200}$, 
respectively for results obtained using the two 
spin priors, $|\chi_{i,z}|< 0.5$ and $|\chi_{i,z}|< 0.9$.
The marginal eccentricity posterior is bounded above by $e_{10}\lesssim0.3$ 
with moderate support for a modest non-zero eccentricity, with the posterior 
maximum near $e_{10}=0.192_{-0.186}^{+0.049}$ 
for the restricted spin prior ($|\chi_{i,z}|< 0.5$) while 
$e_{10}=0.190_{-0.186}^{+0.046}$ for the less restricted spin prior 
($|\chi_{i,z}|< 0.9$). These eccentricity posteriors, while they have a nonzero 
maximum, still have support near zero that may be resolved by 
applying a similar targeting treatment as we did for GW200208\_22.
For both choices of spin prior the posterior on $\chieff$ favors positive 
effective spin, with a maximum at $\chieff=0.285_{-0.124}^{+0.099}$
for the $|\chi_{i,z}|< 0.5$ prior and $\chieff=0.275_{-0.077}^{+0.155}$
for the $|\chi_{i,z}|< 0.9$ prior, which is consistent with 
\cite{RomeroShaw_2021,LIGO-O3-O3a_final-catalog}. For both choices of spin 
prior, the chirp mass posterior has a maximum at 
$\mc = 56.621_{-6.121}^{+4.396} M_{\odot}$ for the spin prior 
$|\chi_{i,z}|< 0.5$ and $\mc = 54.723_{-3.037}^{+7.909} M_{\odot}$
for the spin prior $|\chi_{i,z}|< 0.9$. We obtain parameter values with 
reasonable agreement for mass ratio $q$, detector frame chirp mass $\mc$, 
and effective spin $\chieff$ to the LVK reported values \cite{LIGO-O3-O3a_final-catalog,O3a_PE_data_release}. We provide a summary of the maximum posterior 
values found through a KDE analysis of the posteriors in \figref{fig:190620NRPE} 
in \tabref{tab:190620_kde_params}.

\Figref{fig:190620waveform} shows the highest marginal log-likelihood NR waveform (eGW::35) overlaid with the LIGO Livingston (L1) whitened strain data near GW190620. This best fitting waveform displays close to 3 waveform cycles before merger, which is notably shorter than GW200208\_22, due to GW190620 being about twice as massive. In addition, the GW190620 signal is a better match visually to the best fitting waveform due to the higher SNR of $\sim 12$ as compared to GW200208\_22 with an SNR of $\sim 7$.



\begin{table}
\renewcommand{\arraystretch}{1.5}
  \caption{Summary of Maximum Kernel Density Estimates for the 
  posteriors of GW190620 from \figref{fig:190620NRPE}. 
  }\label{tab:190620_kde_params} 
\begin{ruledtabular}
\begin{tabular}{c|cc}
Parameters  & Fig. 10 NR & Fig. 10 NR  \\
 & $|\chi_{i,z}|< 0.5$ prior & $|\chi_{i,z}|< 0.9$ prior \\ 
\hline 
$\mc (M_\odot)$ & $56.621_{-6.121}^{+4.396}$ & $54.723_{-3.037}^{+7.909}$  \\
$q$ & $0.524_{-0.075}^{+0.182}$ & $0.523_{-0.056}^{+0.200}$  \\
$e_{10}$ & $0.192_{-0.186}^{+0.049}$ & $0.190_{-0.186}^{+0.046}$ \\
$\chieff$ & $0.285_{-0.124}^{+0.099}$ & $0.275_{-0.077}^{+0.155}$ \\
\end{tabular}
\end{ruledtabular}
\end{table}

\section{Conclusions and discussion}
\label{s:conclusion}

In this study, we have systematically investigated the presence of orbital eccentricity in LVK gravitational wave events by combining model-based Bayesian inference with direct comparison to full numerical relativity (NR) simulations. Using the RIFT parameter estimation framework, we \revthree{directly compare} the data for both GW200208\_22 and GW190620 with the eccentric branch of the TEOBResumS waveform model and the extensive RIT catalog of BBH simulations.

The use of the available bank of simulations in the 4th RIT catalog
\cite{RIT_NR4} with numerous eccentric precessing simulations, developed in part for targeting the event GW190521 \cite{Gayathri:2020coq}, 
provides us with the opportunity to assess the possibility of exploring
new eccentric BBH entering the LVK sensitivity band from 10-20 Hz and on.
We have thus applied it to GW200208\_{22} and GW190620 finding a KDE distribution 
favoring eccentricity over quasicircular simulations as displayed in 
Fig.~\ref{fig:200208_kde_ecc} and Fig.~\ref{fig:190620_kde}
(See also recent study on the bias introduced by quasicircular templates on parameter 
estimations \cite{Divyajyoti:2025cwq}). Independently we have used the current 
waveform model \teo{} \cite{teobresums2} to estimate the highest 
likelihood $\lnL$ region in the binary's parameter space. We have covered 
this high likelihood region of parameter space with 42 new simulations 
targeted to GW200208\_{22} searching for 
a new high $\lnL$. We then compared the \teo{} modeled waveforms with precisely 
the corresponding targeted full numerical simulations systematically finding better matching 
for the latter to the GW signal as displayed in Table~\ref{tab:peakLnL}. 
Notably, by also using the 30 recently simulated eccentric waveforms in \cite{Ficarra:2024nro} 
we have found a simulation with an even higher likelihood than any of the 42 targeted 
simulations.

\revthree{It would be very interesting if the case that the targeted 
simulations provide a better match to the GW signals than their corresponding 
models generalizes to other events. Since these NR simulations were chosen to 
target the highest likelihood region of parameter space identified by the 
models, it would be even more interesting if with additional NR simulations 
in their neighborhood (still within the 
90\% \revthree{credible interval}) could produce an
even better match that would yield an even higher $\lnL$.}

Such seems to 
be the case here, since an eccentric simulation, among the 30 nonspinning 
previously performed in \cite{Ficarra:2024nro}, labeled as EccBBH::08, with 
parameters $q=0.4776$ and initial separation $D/M=19.26$ lead to the 
outstanding highest value of almost $\lnL\approx18.7$ (See Table 
\ref{tab:peakLnL}). Notably, for this simulation the initial eccentricity is 
given by 
$e_{1.5PN}=0.233$ (with $e_{20}=0.196$). 
The remnant properties of the 
final merged hole for this EccBBH::08 simulation are given in Table VIII of 
Appendix A of Ref.~\cite{Ficarra:2024nro}.
We also note that in Table~\ref{tab:peakLnL} the top $\lnL$ simulations for 
GW200208\_{22} are eccentric nonprecessing simulations while the precessing 
simulations, eccentric or not, in our catalog fell below, in $\lnL$ terms, as 
can be seen by the respective colorbar maximum values in Fig.~\ref{fig:NRPE2} 
vs. Fig~\ref{fig:NRprecchi1_perp}. 
\revthree{In addition since we see a dependence on the spin priors for moderate spins, an exploration of highly spinning, highly precessing runs might yield further interesting results.}

We applied a parallel methodology to GW190620, utilizing the full breadth of the RIT NR catalog to probe its parameter space. Similar to the results for GW200208\_22, our KDE analysis for GW190620 indicates a preference for non-circular orbits, estimating an eccentricity of $e_{10}=0.190_{-0.186}^{+0.046}$
at a reference frequency of 10\,Hz. While we did not generate a specific targeted set for this event,
table~\ref{tab:190620peakLnL} gives the best matching NR waveform eGW::35 \revthree{(with initial eccentricity 
$e_{1.5PN}=0.244$ (and $e_{10}=0.259$))}, with parameters consistent with the corner plots results in Fig.~\ref{fig:190620NRPE}.  
The consistency of the results with the eccentric hypothesis suggests that GW190620 is another strong candidate
for non-negligible eccentricity. The broad credible intervals remaining in the posterior distributions indicate that GW190620 would likely benefit from the same targeted simulation approach successfully demonstrated here for GW200208\_22 to further constrain its orbital parameters.

The evidence for potential non-negligible eccentricity in both GW200208\_22 and GW190620 points towards formation channels involving dense stellar environments, such as globular clusters or AGN disks, where multi-body interactions can lead to captures and mergers with residual eccentricity. This work establishes a robust pipeline for validating such events.

The success of our purely numerical techniques encourage the targeting of other 
GW events with new simulations in an effort to improve their parameter 
estimates. Among other interesting cases of BBH carrying non negligible 
eccentricity into the LVK sensitivity band to further target studies are GW190620 as well as 
revisiting GW190521 to seek improvements in its parameters. We also note the 
recent study of other potentially eccentric GW signals in \cite{Planas:2025jny} 
identified using phenomenological models and the improvements in 
\texttt{TEOBResumS-Dalí}
models \cite{nagar_2024_teo_dali,Nagar:2024oyk,Albanesi:2025txj}.






\begin{acknowledgements}
This material is based upon work supported by NSF's LIGO Laboratory which is a 
major facility fully funded by the National Science Foundation.
GF gratefully acknowledges the support of University of Calabria through a research fellowship funded by DR 1688/2023.
COL gratefully acknowledges support from NSF awards AST-2319326, PHY-2207920 and PHY-2513442.
ROS gratefully acknowledges support from NSF awards NSF PHY-1912632, 
PHY-2012057, PHY-2309172, AST-2206321, and the Simons Foundation. Finally, the authors thank James Healy for support on using the RIT Catalog of BBH simulations and an anonymous referee for thorough reviews and many valuable suggestions.
\end{acknowledgements}

\appendix

\section{Kernel Density Estimates and Errors}
\label{s:kde}

\begin{figure}[h!tbp]
    \centering
    \includegraphics[width=\columnwidth]{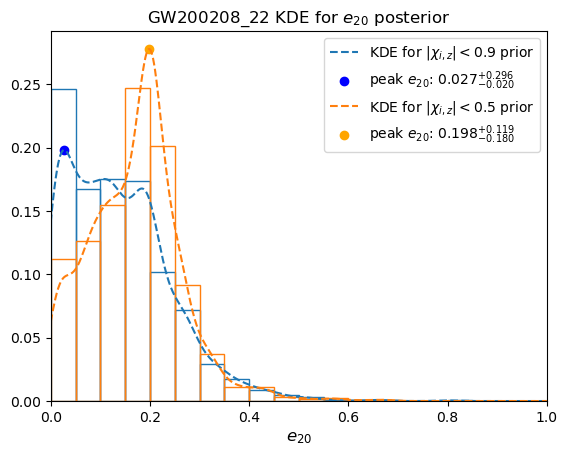}
    \caption{Kernel density estimate of the GW200208\_22 NR-based PE 
    1-dimensional eccentricity posterior at 20\,Hz (histograms from \figref{fig:NRPE2})
    for the two different spin priors, $|\chi_{i,z}|< 0.9$ (blue) and 
    $|\chi_{i,z}|< 0.5$ (orange). The maximum of the KDE for each prior is 
    indicated by a dot of the same color with errors reported as the 90\% \revthree{confidence interval} of the histogram. }
    \label{fig:200208_kde_ecc}
\end{figure}

The 1-dimensional posteriors shown on the diagonal of each corner plot (Figures 
\ref{fig:NRPE2}--\ref{fig:NR_mc20-80},\ref{fig:modelPE1}, and 
\ref{fig:190620NRPE}) can be used as the basis to generate kernel density 
estimates (KDE) for each parameter. We first use the posterior samples that 
RIFT outputs in its second stage (CIP) for each intrinsic parameter, 
$p_{\rm{post}}(\pmb{\lambda})$, to construct a KDE using 
\texttt{scipy.stats.gaussian\_kde} with equally weighted datapoints with weight 
$1/N$ where $N$ is set to the number of posterior samples, 2000. Resultant KDEs 
are evaluated on a grid half the size of the posterior samples over each 
parameter's respective range, $e \in [0,1]$, $q\in[0,1]$, 
$\chi_{\rm{eff}} \in[-1,1]$, and $\mc$ range consistent with the analysis in 
\secref{s:events}, for GW200208\_22 the range is $\mc \in [20,40] \ M_\odot$. 
The maximum of the KDE evaluated over each grid is 
reported as the point with the highest density, this represents the maximum 
parameter value for each intrinsic parameter's 1-dimensional posterior. Errors 
of this maximum value are reported as the 90\% \revthree{confidence interval} of the 
1-dimensional posterior. We note that the maximum parameters as reported from 
the KDE analysis are not the same as the parameters for the peak likelihood 
simulations in \Tabref{tab:peakLnL}. The maximum KDE parameters are determined 
through the analysis of the full posterior over the intrinsic parameters, 
constructed in the CIP stage of RIFT, while the parameters reported in 
\Tabref{tab:peakLnL} are just the intrinsic parameters of the simulation where 
their likelihoods are evaluated in the ILE stage of RIFT. Thus the key 
difference here is the inclusion of the intrinsic prior, the KDE analysis 
of the full intrinsic posterior takes our priors into account while the values 
of \Tabref{tab:peakLnL} do not. As a result we expect a difference between the 
parameters of the peak $\lnL$ simulations in \Tabref{tab:peakLnL} and the 
maximum of the 1-dimensional posteriors for each intrinsic parameter found 
through the KDE analysis. As a demonstration of the method the resultant KDE 
and maximum parameter value is shown overlaid with the 1-dimensional posterior 
histogram for the eccentricity of the NR-based PE of GW200208\_22 in 
\figref{fig:200208_kde_ecc}.

\begin{figure*}[h!tbp]
    \centering
    \includegraphics[width=\columnwidth]{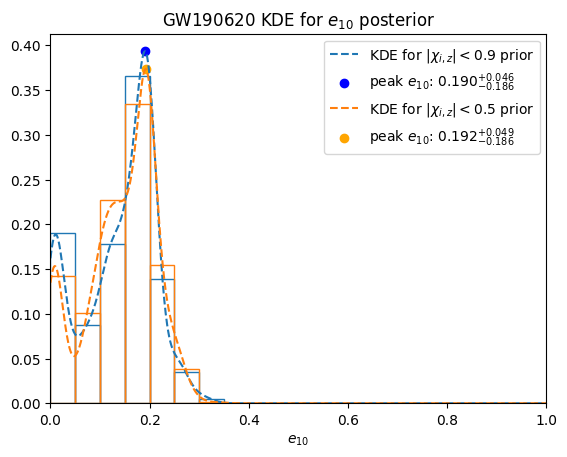}
    \includegraphics[width=\columnwidth]{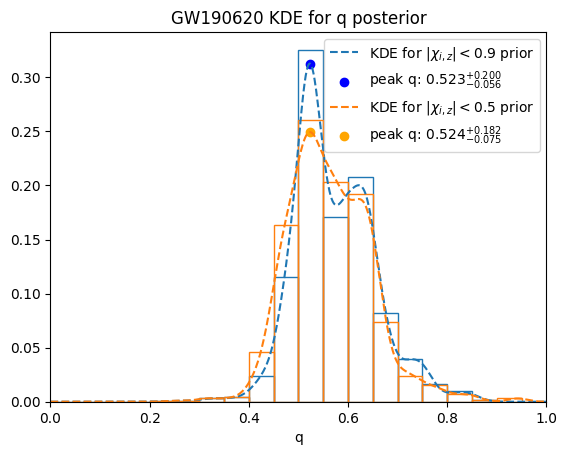}
    \includegraphics[width=\columnwidth]{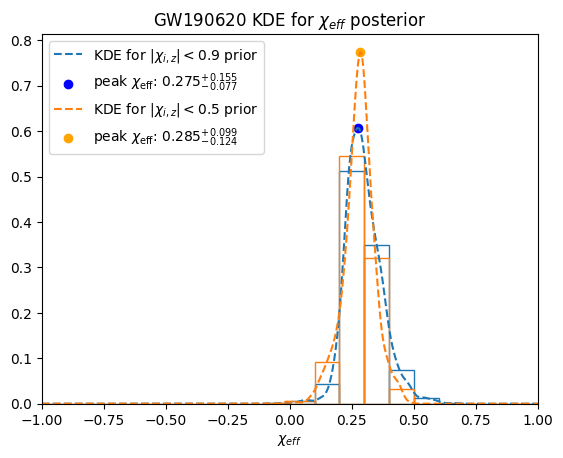}
    \includegraphics[width=\columnwidth]{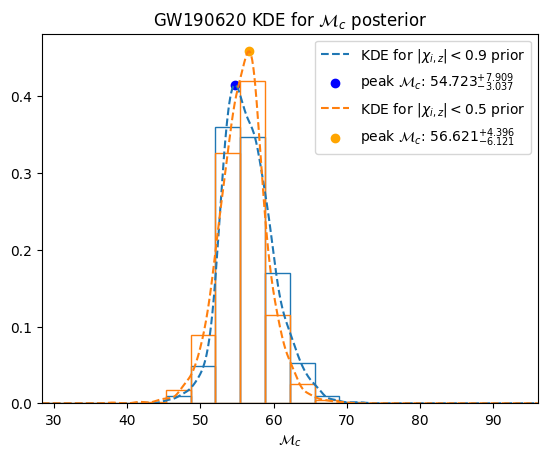}
    \caption{Kernel density estimate of the GW190620 NR-based PE 
    1-dimensional posteriors for eccentricity at 10\,Hz (top left), mass ratio $q$ 
    (top right), effective spin $\chieff$ (bottom left), and chirp mass $\mc$ 
    (bottom right) (histograms from \figref{fig:190620NRPE})
    for the two different spin priors, $|\chi_{i,z}|< 0.9$ (blue) and 
    $|\chi_{i,z}|< 0.5$ (orange). The maximum of the KDE for each prior is 
    indicated by a dot of the same color with errors reported as the 90\% \revthree{confidence interval} of the histogram.}
    \label{fig:190620_kde}
\end{figure*}

We asses the errors of the maximum parameter value by generating 1000 
additional posterior realizations, where the real posterior samples are 
resampled by choosing random rows from the real posterior samples with 
replacement such that our resampled posterior is the same size as the real 
posterior samples. For each posterior sample realization the KDE is computed
then evaluated on the same grid as used for the real posterior sample, and the 
same maximum finding process is applied. The standard deviation and the 
symmetric 90\% \revthree{confidence interval} on the set of maximum parameter values is 
computed where these errors represent the statistical error for the maximum 
parameter value obtained through the KDE. For instance, the standard deviation 
of the location of the maximum eccentricity found in 
\figref{fig:200208_kde_ecc} are $0.029 \pm 0.014$ for the 
$|\chi_{i,z}|< 0.9$ prior and $0.198\pm0.001$ for the 
$|\chi_{i,z}|< 0.5$ prior. The 90\% \revthree{confidence interval} on the set of maximum 
eccentricity values in \figref{fig:200208_kde_ecc} are 
$0.029_{-0.005}^{+0.003}$ for the $|\chi_{i,z}|< 0.9$ prior and 
$ 0.198_{-0.002}^{+0.002}$ for the $|\chi_{i,z}|< 0.5$ prior. 
Similarly, errors on the other maximum intrinsic parameters ($\chieff, \ \mc, 
\ q$, etc.) are on the same order of magnitude, which indicates that the 
statistical error in the KDE maximum finding method is negligible.

We show in \figref{fig:190620_kde} the KDE for each parameter of GW190620, 
in this case the two spin priors yield very consistent results returning nearly 
the same values for each spin prior.

\section{NR simulation parameters}\label{s:NRsimsparam}

\CarlosO{Here we provide the details of the 42 targeted simulations specifically
performed for the studies in this paper (where we
took the opportunity to experiment with the
slow-start-lapse gauge \cite{Etienne:2024ncu})}, in addition to the
1881 simulations in the fourth RIT catalog \cite{RIT_NR4} and
the additional 30 simulations in \cite{Ficarra:2024nro}.
In these cases a small $P_r$ component has been added to the initial 
quasicircular parameters from the instantaneous $P_t$ up to 3.5PN 
radiation terms \cite{Healy:2017zqj,Ciarfella:2024clj}
in order to ensure a lower eccentricity.
In all other explicit eccentric cases, as those given here in Table~\ref{tab:eGWidparameters}, the prescription
consists on setting $P_r=0$ \cite{Ciarfella:2022hfy}.
\begin{table*}
\caption{Initial data parameters for the sequence of simulations described in Sec.\ref{ss:NRsims} with a larger black hole (labeled 1) and a smaller black hole (labeled 2). Punctures are located at $\Vec{r}_1=(x_1,0,0)$ and $\Vec{r}_2=(x_2,0,0)$ with mass ratio $q=m^H_2/m^H_1$, eccentricity parameter $f$, linear momenta $\vec{P}=\pm(0,P_t,0)$, puncture masses $m^p/M$, horizon (Christodoulou) masses $m^H/M$, dimensionless spin parameter $\chi$ and total ADM mass $M_{\rm{ADM}}/M$. We also provide initial values of the Newtonian eccentricity $e_0 = 2f-f^2$ and PN orders estimates. 
}
\begin{tabular}{lccccccccccccccc}
\toprule
Run & $x_1/M$ & $x_2/M$ & $q$ & $f$ & $P_t/M$ & $m^p_1/M$ & $m^p_2/M$ & $m^H_1/M$ & $m^H_2/M$ & $\chi_{1}$ & $\chi_{2}$ & $M_{\rm{ADM}}/M$ & $e_0$ & $e_{1.5\rm{PN}}$ & $e_{3.5\rm{PN}}$ \\
\hline
eGW::01 & 4.12 & -8.88 & 0.465 & 0.075 & 0.0641 & 0.6567 & 0.3078 & 0.6826 & 0.3174 & 0.2482 & 0.0335 & 0.9915 & 0.144 & 0.187 & 0.207 \\
eGW::02 & 4.16 & -8.84 & 0.470 & 0.075 & 0.0650 & 0.6706 & 0.3101 & 0.6801 & 0.3199 & 0.0483 & -0.0506 & 0.9915 & 0.144 &  0.191 & 0.216 \\
eGW::03 & 4.40 & -8.60 & 0.512 & 0.075 & 0.0663 & 0.6290 & 0.3288 & 0.6613 & 0.3387 & 0.2926 & -0.0256 & 0.9912 & 0.144 & 0.187 & 0.206 \\
eGW::04 & 4.45 & -8.55 & 0.520 & 0.075 & 0.0664 & 0.6361 & 0.3205 & 0.6579 & 0.3421 & 0.2180 & 0.2930 & 0.9912 & 0.144 &  0.185 & 0.203 \\
eGW::05 & 4.46 & -8.54 & 0.523 & 0.075 & 0.0667 & 0.6323 & 0.3313 & 0.6564 & 0.3436 & 0.2367 & 0.1360 & 0.9912 & 0.144 & 0.186 & 0.205 \\
eGW::06 & 4.46 & -8.54 & 0.524 & 0.075 & 0.0666 & 0.6147 & 0.3339 & 0.6563 & 0.3437 & 0.3445 & -0.0214 & 0.9911 & 0.144 & 0.186 & 0.204 \\
eGW::07 & 4.63 & -8.37 & 0.553 & 0.075 & 0.0684 & 0.6343 & 0.3439 & 0.6437 & 0.3563 & -0.0007 & 0.1314 & 0.9910 & 0.144 & 0.190 & 0.213 \\
eGW::08 & 4.62 & -8.38 & 0.553 & 0.075 & 0.0683 & 0.6137 & 0.3296 & 0.6439 & 0.3561 & 0.2812 & -0.3379 & 0.9911 & 0.144 & 0.190 & 0.212 \\
eGW::09 & 4.71 & -8.29 & 0.570 & 0.075 & 0.0686 & 0.6051 & 0.3481 & 0.6371 & 0.3629 & 0.2942 & -0.1823 & 0.9910 & 0.144 & 0.188 & 0.208 \\
eGW::10 & 4.76 & -8.24 & 0.579 & 0.075 & 0.0694 & 0.6227 & 0.3531 & 0.6334 & 0.3666 & 0.0678 & -0.1531 & 0.9909 & 0.144 & 0.192 & 0.216 \\
eGW::11 & 4.80 & -8.20 & 0.586 & 0.075 & 0.0688 & 0.5927 & 0.3591 & 0.6304 & 0.3696 & 0.3287 & 0.0573 & 0.9909 & 0.144 & 0.185 & 0.203 \\
eGW::12 & 4.86 & -8.14 & 0.599 & 0.075 & 0.0694 & 0.5850 & 0.3622 & 0.6255 & 0.3745 & 0.3448 & -0.1223 & 0.9908 & 0.144 & 0.187 & 0.206 \\
eGW::13 & 4.88 & -8.12 & 0.602 & 0.075 & 0.0694 & 0.5932 & 0.3655 & 0.6243 & 0.3757 & 0.2895 & 0.0222 & 0.9908 & 0.144 & 0.186 & 0.205 \\
eGW::14 & 4.95 & -8.05 & 0.617 & 0.075 & 0.0696 & 0.5762 & 0.3709 & 0.6184 & 0.3816 & 0.3557 & 0.0560 & 0.9907 & 0.144 & 0.185 & 0.202 \\
eGW::15 & 4.97 & -8.03 & 0.619 & 0.075 & 0.0700 & 0.5967 & 0.3705 & 0.6175 & 0.3825 & 0.2108 & 0.0759 & 0.9907 & 0.144 & 0.187 & 0.207 \\
eGW::16 & 5.01 & -7.99 & 0.629 & 0.075 & 0.0700 & 0.5711 & 0.3758 & 0.6140 & 0.3860 & 0.3599 & 0.0042 & 0.9907 & 0.144 & 0.186 & 0.203 \\
eGW::17 & 5.03 & -7.97 & 0.632 & 0.075 & 0.0701 & 0.5845 & 0.3724 & 0.6128 & 0.3872 & 0.2730 & 0.1722 & 0.9907 & 0.144 & 0.186 & 0.203 \\
eGW::18 & 5.05 & -7.95 & 0.635 & 0.075 & 0.0703 & 0.5782 & 0.3783 & 0.6114 & 0.3886 & 0.3059 & 0.0048 & 0.9907 & 0.144 & 0.187 & 0.205 \\
eGW::19 & 5.06 & -7.94 & 0.637 & 0.075 & 0.0703 & 0.5776 & 0.3784 & 0.6107 & 0.3893 & 0.3048 & 0.0639 & 0.9907 & 0.144 & 0.186 & 0.204 \\
eGW::20 & 5.06 & -7.94 & 0.638 & 0.075 & 0.0701 & 0.5604 & 0.3787 & 0.6106 & 0.3894 & 0.3962 & 0.0584 & 0.9907 & 0.144 & 0.184 & 0.201 \\
eGW::21 & 5.20 & -7.80 & 0.668 & 0.075 & 0.0710 & 0.5607 & 0.3902 & 0.5994 & 0.4006 & 0.3407 & 0.0286 & 0.9906 & 0.144 & 0.185 & 0.203 \\
eGW::22 & 5.54 & -7.46 & 0.744 & 0.075 & 0.0723 & 0.5556 & 0.4034 & 0.5735 & 0.4265 & 0.1841 & 0.2711 & 0.9904 & 0.144 & 0.186 & 0.203 \\
eGW::23 & 5.66 & -7.34 & 0.772 & 0.075 & 0.0728 & 0.5374 & 0.4250 & 0.5644 & 0.4356 & 0.2708 & 0.0186 & 0.9904 & 0.144 & 0.187 & 0.206 \\
eGW::24 & 4.33 & -8.67 & 0.500 & 0.050  & 0.0680 & 0.6548 & 0.3236 & 0.6667 & 0.3333 & 0.1000 & 0.0000 & 0.9917 & 0.098 & 0.128 & 0.141 \\
eGW::25 & 4.33 & -8.67 & 0.500 & 0.050  & 0.0679 & 0.6515 & 0.3236 & 0.6667 & 0.3333 & 0.1500 & 0.0000 & 0.9917 & 0.098 & 0.127 & 0.140 \\
eGW::26 & 4.33 & -8.67 & 0.500 & 0.050  & 0.0677 & 0.6468 & 0.3236 & 0.6667 & 0.3333 & 0.2000 & 0.0000 & 0.9917 & 0.098 & 0.126 & 0.139 \\
eGW::27 & 4.33 & -8.67 & 0.500 & 0.050  & 0.0675 & 0.6329 & 0.3236 & 0.6667 & 0.3333 & 0.3000 & 0.0000 & 0.9917 & 0.098  & 0.125 & 0.137 \\
eGW::28 & 4.33 & -8.67 & 0.500 & 0.050  & 0.0673 & 0.6124 & 0.3236 & 0.6667 & 0.3333 & 0.4000 & 0.0000 & 0.9917 & 0.098 & 0.124 & 0.135 \\
eGW::29 & 4.33 & -8.67 & 0.500 & 0.050  & 0.0671 & 0.5993 & 0.3236 & 0.6667 & 0.3333 & 0.4500 & 0.0000 & 0.9917 & 0.098 & 0.123 & 0.134 \\
eGW::30 & 4.32 & -8.68 & 0.500 & 0.050  & 0.0668 & 0.5450 & 0.3236 & 0.6667 & 0.3333 & 0.6000 & 0.0000 & 0.9917 & 0.098 & 0.122 & 0.131 \\
eGW::31 & 4.33 & -8.67 & 0.500 & 0.100 & 0.0644 & 0.6549 & 0.3237 & 0.6667 & 0.3333 & 0.1000 & 0.0000 & 0.9909 & 0.190 & 0.251 & 0.285 \\
eGW::32 & 4.33 & -8.67 & 0.500 & 0.100  & 0.0643 & 0.6516 & 0.3237 & 0.6667 & 0.3333 & 0.1500 & 0.0000 & 0.9909 & 0.190 & 0.250 & 0.282 \\
eGW::33 & 4.33 & -8.67 & 0.500 & 0.100  & 0.0642 & 0.6469 & 0.3237 & 0.6667 & 0.3333 & 0.2000 & 0.0000 & 0.9909 & 0.190 & 0.248 & 0.279 \\
eGW::34 & 4.33 & -8.67 & 0.500 & 0.100  & 0.0639 & 0.6330 & 0.3237 & 0.6667 & 0.3333 & 0.3000 & 0.0000 & 0.9909 & 0.190 & 0.246 & 0.274 \\
eGW::35 & 4.32 & -8.68 & 0.500 & 0.100  & 0.0637 & 0.6125 & 0.3238 & 0.6667 & 0.3333 & 0.4000 & 0.0000 & 0.9909 & 0.190 & 0.244 & 0.269 \\
eGW::36 & 4.32 & -8.68 & 0.500 & 0.100 & 0.0636 & 0.5994 & 0.3238 & 0.6667 & 0.3333 & 0.4500 & 0.0000 & 0.9909 & 0.190 & 0.242 & 0.267 \\
eGW::37 & 4.32 & -8.68 & 0.500 & 0.100 & 0.0633 & 0.5451 & 0.3238 & 0.6667 & 0.3333 & 0.6000 & 0.0000 & 0.9908 & 0.190 & 0.239 & 0.260 \\
eGW::38 & 4.87 & -8.13 & 0.600 & 0.025 & 0.0731 & 0.5889 & 0.3647 & 0.6250 & 0.3750 & 0.3200 & 0.0000 & 0.9917 & 0.049 & 0.063 & 0.068 \\
eGW::39 & 5.35 & -7.65 & 0.700 & 0.025 & 0.0755 & 0.5500 & 0.4012 & 0.5882 & 0.4118 & 0.3400 & 0.0000 & 0.9914 & 0.049 & 0.063 & 0.068 \\
\addition{eGW::40} & 5.97 & -7.03 & 0.850 & 0.175 & 0.0662 & 0.5303 & 0.4491 & 0.5405 & 0.4595 & 0.0000 & 0.0000 & 0.9885 & 0.319 & 0.431 & 0.479 \\
\addition{eGW::41} & 5.98 & -7.02 & 0.850 & 0.175 & 0.0669 & 0.4996 & 0.4491 & 0.5405 & 0.4595 & -0.3700 & 0.0000 & 0.9886 & 0.319 & 0.444 & 0.497 \\
\addition{eGW::42} & 5.97 & -7.03 & 0.850 & 0.175 & 0.0655 & 0.4996 & 0.4492 & 0.5405 & 0.4595 & 0.3700 & 0.0000 & 0.9885 & 0.319 & 0.418 & 0.460 \\
\hline
\end{tabular}
\label{tab:eGWidparameters}
\end{table*}

In Table~\ref{tab:eGWfinalproperties} we provide the merged black
hole properties, final mass, spin and recoil velocity of the remnant
hole. We also provide a measure of the radiated gravitational energy,
consistent with the final mass of the merged hole and waveform properties
such as the peak luminosity, amplitude and frequency of the leading
(2,2)-mode. The time and number of orbits to merger of our simulations
completed the table. These properties are given in the format of the
metadata in the RIT BBH waveforms catalog
\url{https://ccrgpages.rit.edu/~RITCatalog/}.
\begin{table*}
\caption{Properties of the sequence of simulations described in Sec.\ref{ss:NRsims}. We report the remnant mass $M_f/M$ and spin $\chi_f$, the radiated energy $\delta\mathcal{M}=M_{\rm{ADM}}-M_f$, merger time $t_{\rm{m}}/M$, number of orbits $N$, strain peak amplitude $\left(r/M\right)|h_{22}^{\rm{peak}}|$, recoil velocity $V_{\rm{kick}}$, peak frequency $M\omega_{22}^{\rm{peak}}$ and peak luminosity $\mathcal{L}_{\rm{peak}}$.}
\begin{tabular}{lccccccccc}
\toprule
Run & $M_f/M$ & $\chi_f$ & $\delta\mathcal{M}/M$ & $t_{\rm{m}}/M$ & $N$ & $V_{\rm{kick}}$[km/s] & $\left(r/M\right)|h_{22}^{\rm{peak}}|$ & $M\omega_{22}^{\rm{peak}}$ & $\mathcal{L}_{\rm{peak}}$[$10^{-56}$erg/s] \\
\hline
eGW::01 & 0.9574 & 0.6946 & 0.0341 & 1291.5 & 8.22 & 97.21 & 0.3355 & 0.3594 & 2.9193 \\
eGW::02 & 0.9620 & 0.6241 & 0.0295 & 1115.6 & 7.17 & 138.34 & 0.3337 & 0.3457 & 2.6316 \\
eGW::03 & 0.9543 & 0.7157 & 0.0370 & 1267.5 & 8.11 & 76.37 & 0.3492 & 0.3663 & 3.2155 \\
eGW::04 & 0.9538 & 0.7125 & 0.0374 & 1302.8 & 8.34 & 107.35 & 0.3479 & 0.3660 & 3.1797 \\
eGW::05 & 0.9539 & 0.7105 & 0.0372 & 1267.9 & 8.11 & 92.05 & 0.3518 & 0.3659 & 3.2322 \\
eGW::06 & 0.9528 & 0.7320 & 0.0384 & 1295.2 & 8.31 & 65.62 & 0.3489 & 0.3684 & 3.2745 \\
eGW::07 & 0.9580 & 0.6466 & 0.0331 & 1093.1 & 7.04 & 143.03 & 0.3542 & 0.3512 & 2.9670 \\
eGW::08 & 0.9562 & 0.6973 & 0.0349 & 1140.4 & 7.38 & 76.56 & 0.3551 & 0.3596 & 3.1904 \\
eGW::09 & 0.9539 & 0.7153 & 0.0371 & 1187.2 & 7.65 & 67.23 & 0.3627 & 0.3648 & 3.4162 \\
eGW::10 & 0.9507 & 0.7347 & 0.0401 & 1216.9 & 7.81 & 58.00 & 0.3692 & 0.3708 & 3.6142 \\
eGW::11 & 0.9498 & 0.7402 & 0.0410 & 1266.8 & 8.13 & 57.15 & 0.3635 & 0.3714 & 3.5460 \\
eGW::12 & 0.9572 & 0.6541 & 0.0338 & 1041.8 & 6.70 & 101.51 & 0.3639 & 0.3496 & 3.1338 \\
eGW::13 & 0.9506 & 0.7309 & 0.0402 & 1225.9 & 7.88 & 59.30 & 0.3695 & 0.3707 & 3.6047 \\
eGW::14 & 0.9483 & 0.7494 & 0.0425 & 1266.8 & 8.15 & 49.12 & 0.3682 & 0.3733 & 3.6674 \\
eGW::15 & 0.9514 & 0.7202 & 0.0394 & 1199.6 & 7.70 & 73.55 & 0.3724 & 0.3663 & 3.5945 \\
eGW::16 & 0.9482 & 0.7479 & 0.0425 & 1247.6 & 8.03 & 49.62 & 0.3718 & 0.3729 & 3.7094 \\
eGW::17 & 0.9485 & 0.7399 & 0.0422 & 1248.4 & 8.02 & 61.94 & 0.3722 & 0.3718 & 3.6762 \\
eGW::18 & 0.9494 & 0.7368 & 0.0413 & 1215.0 & 7.82 & 51.03 & 0.3749 & 0.3715 & 3.7142 \\
eGW::19 & 0.9486 & 0.7406 & 0.0420 & 1231.1 & 7.91 & 51.36 & 0.3746 & 0.3725 & 3.7327 \\
eGW::20 & 0.9469 & 0.7604 & 0.0437 & 1282.1 & 8.26 & 44.67 & 0.3695 & 0.3751 & 3.7412 \\
eGW::21 & 0.9474 & 0.7480 & 0.0432 & 1227.8 & 7.90 & 44.79 & 0.3788 & 0.3746 & 3.8399 \\
eGW::22 & 0.9466 & 0.7422 & 0.0438 & 1206.8 & 7.78 & 66.17 & 0.3873 & 0.3727 & 3.9463 \\
eGW::23 & 0.9483 & 0.7361 & 0.0421 & 1158.8 & 7.50 & 35.60 & 0.3876 & 0.3698 & 3.8918 \\
eGW::24 & 0.9592 & 0.6563 & 0.0325 & 1616.2 & 9.28 & 124.28 & 0.3469 & 0.3534 & 2.9266 \\
eGW::25 & 0.9580 & 0.6710 & 0.0337 & 1651.5 & 9.49 & 113.40 & 0.3457 & 0.3547 & 2.9632 \\
eGW::26 & 0.9572 & 0.6857 & 0.0345 & 1689.4 & 9.74 & 100.06 & 0.3435 & 0.3580 & 2.9887 \\
eGW::27 & 0.9552 & 0.7185 & 0.0365 & 1770.0 & 10.23 & 78.68 & 0.3453 & 0.3661 & 3.1732 \\
eGW::28 & 0.9521 & 0.7487 & 0.0396 & 1845.0 & 10.66 & 63.72 & 0.3466 & 0.3733 & 3.3564 \\
eGW::29 & 0.9508 & 0.7630 & 0.0409 & 1883.2 & 10.91 & 57.21 & 0.3445 & 0.3764 & 3.3939 \\
eGW::30 & 0.9458 & 0.8095 & 0.0459 & 2002.9 & 11.67 & 54.34 & 0.3478 & 0.3948 & 3.7727 \\
eGW::31 & 0.9595 & 0.6523 & 0.0314 & 790.1 & 5.74 & 124.18 & 0.3396 & 0.3502 & 2.7935 \\
eGW::32 & 0.9591 & 0.6704 & 0.0318 & 828.4 & 5.98 & 113.27 & 0.3414 & 0.3577 & 2.8920 \\
eGW::33 & 0.9579 & 0.6888 & 0.0330 & 863.2 & 6.18 & 95.92 & 0.3466 & 0.3598 & 3.0655 \\
eGW::34 & 0.9543 & 0.7191 & 0.0366 & 921.0 & 6.53 & 83.00 & 0.3493 & 0.3639 & 3.2444 \\
eGW::35 & 0.9524 & 0.7463 & 0.0384 & 984.0 & 7.02 & 58.36 & 0.3413 & 0.3716 & 3.2240 \\
eGW::36 & 0.9517 & 0.7633 & 0.0392 & 1020.4 & 7.27 & 57.75 & 0.3419 & 0.3781 & 3.3437 \\
eGW::37 & 0.9451 & 0.8085 & 0.0458 & 1119.0 & 7.92 & 55.30 & 0.3489 & 0.3890 & 3.7663 \\
eGW::38 & 0.9504 & 0.7364 & 0.0413 & 2266.9 & 11.91 & 55.95 & 0.3671 & 0.3705 & 3.5854 \\
eGW::39 & 0.9477 & 0.7465 & 0.0437 & 2205.0 & 11.62 & 44.90 & 0.3808 & 0.3743 & 3.8539 \\
\addition{eGW::40} & 0.9492 & 0.6914 & 0.03938 & 175.1 & 1.78 & 47.26 & 0.4184 & 0.3576 & 4.1337 \\
\addition{eGW::41} & 0.9587 & 0.6343 & 0.02988 & 154.5 & 1.38 & 140.64 & 0.4171 & 0.3421 & 3.8928 \\
\addition{eGW::42} & 0.9483 & 0.7504 & 0.04016 & 269.2 & 2.88 & 66.00 & 0.3817 & 0.3756 & 3.8800 \\
\hline
\end{tabular}
\label{tab:eGWfinalproperties}
\end{table*}

\newpage
\bibliography{biblio,LIGO-publications,extra,references}

\end{document}